\documentclass[acmtosem,screen]{acmart}

\usepackage[utf8]{inputenc}
\usepackage[english]{babel}

\usepackage{graphicx}
\DeclareGraphicsExtensions{.pdf,.png,.jpg}

\usepackage{booktabs}
\usepackage{multirow}
\usepackage{enumitem}
\usepackage{makecell}
\usepackage{relsize}

\usepackage[dvipsnames]{xcolor}
\usepackage[most]{tcolorbox}
\usepackage[normalem]{ulem}
\usepackage{xcolor,colortbl}

\definecolor{darkgreen}{RGB}{0,100,0}
\definecolor{codegreen}{rgb}{0,0.5,0}
\definecolor{codepurple}{rgb}{0.58,0,0.82}
\definecolor{codegray}{rgb}{0.5,0.5,0.5}
\usepackage{listings}
\lstdefinestyle{mystyle}{
  commentstyle=\color{codegreen},
  keywordstyle=\bfseries,
  stringstyle=\color{codepurple},
  basicstyle=\ttfamily\scriptsize,
  breaklines=true,
  captionpos=b,
  keepspaces=true,
  tabsize=2
}
\usepackage[noend,ruled,linesnumbered]{algorithm2e}

\usepackage{tikz}
\usetikzlibrary{arrows.meta,positioning,shapes,calc} 

\usepackage{xcolor}
\usepackage{xparse} 

\newcommand{\find}[1]{%
\begin{tcolorbox}[tile,size=fbox,boxsep=2mm,boxrule=0pt,top=0pt,bottom=0pt,
borderline={0.6mm}{0pt}{black!66!white},colback=black!5!white]
\em #1
\end{tcolorbox}
}

\newboolean{COMMENTSON} 
\setboolean{COMMENTSON}{true}   
\ifthenelse{\boolean{COMMENTSON}}
{

}

\definecolor{DarkOrange}{rgb}{0.8,0.3,0.0} 
\definecolor{DarkCyan}{rgb}{0.0, 0.55, 0.55}
\definecolor{codegreen}{rgb}{0,0.6,0}
\definecolor{codegray}{rgb}{0.5,0.5,0.5}
\definecolor{codepurple}{rgb}{0.58,0,0.82}
\definecolor{backcolour}{rgb}{0.95,0.95,0.92}

\newcommand{\papertitle}{Dissecting Software Graphs: Structural Insights for Driver-Guided Fuzzing}
\newcommand{\papertitleshort}{Dissecting Software Graphs: Structural Insights for Driver-Guided Fuzzing}

\newcommand{\mfuzz}{\mbox{\textsc{MFuzz}}}
\newcommand{\graphdist}{\mbox{\textsc{GDissert}}}

\newcommand{\paperkeywords}{Fuzzing, Program Analysis, Software Security}

\newcommand{\authorAname}{Wen Li}
\newcommand{\authorAaffil}{Utah State University}
\newcommand{\authorAemail}{awen.li@usu.edu}

\newcommand{\authorBname}{Baihong Chen}
\newcommand{\authorBaffil}{Utah State University}
\newcommand{\authorBemail}{b.chen@usu.edu}

\newcommand{\authorCname}{Hua Ming}
\newcommand{\authorCaffil}{University of Michigan}
\newcommand{\authorCemail}{huaming@umich.edu}

\newcommand{\authorDname}{Weifeng Pan}
\newcommand{\authorDaffil}{Zhejiang Gongshang University}
\newcommand{\authorDemail}{wfpan@zjgsu.edu.cn}

\newcommand{\authorEname}{Haipeng Cai}
\newcommand{\authorEaffil}{University at Buffalo, SUNY}
\newcommand{\authorEemail}{haipengc@buffalo.edu}

\newcommand{\authorFname}{Tian Xie}
\newcommand{\authorFaffil}{Utah State University}
\newcommand{\authorFemail}{tian.xie@usu.edu}

\newcommand{\appendixcontent}{%
  \appendix
}

\newcommand{\blackcircleone}[1]{%
  \tikz[baseline=(char.base)]{
    \node[shape=circle, fill=black, text=white, inner sep=1pt] (char) {\small #1};
  }%
}

\setcopyright{acmcopyright}

\begin{document}

\title[\papertitleshort]{\papertitle}

\author{\authorBname}
\affiliation{%
  \institution{\authorBaffil}
  \country{USA}
}
\email{\authorBemail}
\authornote{First author}

\author{\authorCname}
\affiliation{%
  \institution{\authorCaffil}
  \country{USA}
}
\email{\authorCemail}

\author{\authorDname}
\affiliation{%
  \institution{\authorDaffil}
  \country{China}
}

\email{\authorDemail}

\author{\authorFname}
\affiliation{%
  \institution{\authorFaffil}
  \country{USA}
}
\email{\authorFemail}

\author{\authorEname}
\affiliation{%
  \institution{\authorEaffil}
  \country{USA}
}
\email{\authorEemail}

\author{\authorAname}
\affiliation{%
  \institution{\authorAaffil}
  \country{USA}
}
\email{\authorAemail}
\authornote{Corresponding author}

\ccsdesc[500]{Software and its engineering~Software testing and debugging}
\ccsdesc[500]{Security and privacy~Software and application security}

\begin{abstract}

Many software systems expose multiple execution modes through command-line options, subcommands, and configuration flags. For such programs, fuzzing depends on both mutated inputs and the invoked mode. Yet evaluations still focus on coverage and bug counts, leaving unclear how execution modes partition, overlap, and miss software structure, and how these differences affect effectiveness.
We present an empirical study of software structure under multi-driver fuzzing. We propose a structural abstraction that uses a static call graph as a shared backbone and projects driver-specific dynamic coverage onto it to derive driver-induced subgraphs. Based on this abstraction, we develop a four-phase methodology for backbone construction, fuzzing and profiling, graph-based analysis, and research-question-driven evaluation. We apply it to 27 OSS-Fuzz-derived C/C++ projects, spanning 43 executables and 854 driver configurations.

Under the same total budget, multi-driver fuzzing outperforms the best single-driver baseline, increasing covered call-graph nodes by 27.9\% and CFG-edge coverage by 73.5\%, and revealing 11 unique bugs and abnormal behaviors largely missed by single-driver fuzzing. However, driver contributions are uneven, subgraphs differ substantially in cohesion, fragmentation, modularity, overlap, and residual under-exploration follows recurring regimes rather than a homogeneous tail. These results show that multi-driver fuzzing is fundamentally a structural exploration problem.

\end{abstract}

\keywords{\paperkeywords}

\maketitle

\section{Introduction}
\label{sec:intro}

Modern software systems often expose rich functionality 
through command-line options, subcommands, configuration flags, and input formats.
In such programs, 
different execution modes can activate substantially different parsing logic, control-flow paths, 
and downstream functionality.
For fuzzing, 
this means that software is rarely explored through a single uniform mode.
Instead, 
the reachable behavior of a program is shaped jointly by mutated inputs and 
by the execution mode under which the program is invoked.
This challenge is especially relevant for command-line and 
configuration-rich software, 
where invocation choices can partition program behavior 
into partially overlapping functional regions~\cite{schroder2022empirical,li2018fuzzing,afl-fuzz,swiecki2016honggfuzz}.

Existing fuzzing research has largely focused on improving testing effectiveness through better guidance, mutation strategies, static analysis, and hybrid analysis, 
including work on command-line and configuration-rich software as well as broader program-environment fuzzing~\cite{zhu2022fuzzing,li2023polyfuzz,zhang2020survey,wang2023carpetfuzz,lu2024fuzzing,lee2025zigzagfuzz,Meng2024EnvFuzz}. These efforts have significantly advanced fuzzing, 
while empirical studies have also improved understanding of 
how fuzzers should be evaluated, 
typically through aggregate outcomes such as code coverage, bug-finding effectiveness, crashes, 
and benchmark quality~\cite{klees2018evaluating,fuzzbench,magma,ossfuzz}. 
However, 
prior work has primarily treated options, execution modes, and environmental diversity as means to improve fuzzing effectiveness, rather than as objects of structural analysis. 
As a result, 
they reveal relatively little about how different execution modes of the same program relate 
to one another structurally. 
In particular, 
it remains poorly understood whether multiple execution modes explore complementary functional regions, 
whether they are largely redundant, how their induced execution footprints are organized, 
and why substantial parts of a program may remain under-explored even under multi-driver fuzzing. 
This gap leaves driver complementarity, 
redundancy, and residual blind spots insufficiently characterized, 
with potential implications for fuzzing efficiency under multi-driver settings.

This paper takes a different perspective. Rather than proposing a new fuzzing
algorithm, it presents an empirical study of software structure as observed
through multi-driver fuzzing. Our key idea is to analyze different execution
modes in a shared graph space: for each target program, we use a static
function-level call graph as a common structural backbone and map driver-specific
dynamic coverage onto that backbone to derive driver-induced subgraphs. This
representation makes it possible to compare how execution modes partition,
share, and miss program structure while keeping the analysis grounded in
behavior that fuzzing actually exercises.
Using this perspective, we design a four-phase empirical methodology that
combines structural backbone construction, multi-driver fuzzing and profiling,
graph-based structural analysis, and research-question-driven evaluation. We
implement this methodology with lightweight execution and analysis support, and
apply it to a diverse benchmark suite derived from OSS-Fuzz, covering 27
projects, 43 executables, and 854 driver configurations across multiple
software domains,
including media processing, document and parser utilities, 
toolchain and binary analysis tools, 
language runtimes, archive and compression software, network/protocol utilities, and database/storage systems.

Our study shows that the effectiveness of multi-driver fuzzing is fundamentally shaped by software structure.
The key findings and implications are summarized as follows:
\begin{itemize}
    \item Multi-driver fuzzing generally outperforms the best single-driver baseline under the same total budget, improving coverage by an average of 27.9\% in covered call-graph nodes and 73.5\% in covered CFG edges. Beyond coverage, {\mfuzz} also demonstrates clear practical value for bug discovery: it exposed 11 unique bugs and abnormal behaviors across the benchmark suite, most of which were not recovered by the strongest single-driver configuration. These results indicate that preserving driver diversity is important not only for expanding structural reach, but also for uncovering semantically specialized failure behaviors that remain inaccessible to single-driver fuzzing.
    
    \item Equal-budget round-robin execution yields highly uneven driver contributions, and driver-induced subgraphs differ substantially in size, cohesion, fragmentation, modularity, and pairwise overlap. This shows that multi-driver fuzzing is not simply a matter of adding more drivers: its benefit depends on the marginal structural contribution, complementarity, and redundancy of drivers.
    
    \item The remaining cold structure does not form a single homogeneous tail. Instead, executables exhibit recurring residual-gap regimes, ranging from backbone-dominated residual gaps, to substantial residual tails, to localized residual pockets. This suggests that persistent under-exploration takes qualitatively different forms across executables.
    
    \item These findings suggest several practical directions for driver-aware fuzzing. Effective approaches should measure the marginal value of each driver, account for overlap and redundancy among drivers, prioritize drivers that continue to extend exploration into weakly covered regions, and adapt effort allocation to different residual regimes, rather than relying solely on uniform round-robin execution or aggregate coverage.
\end{itemize}

Taken together, 
these results show that multi-driver fuzzing is not merely a matter of using more drivers, but of understanding how different execution modes partition, overlap, and leave under-explored regions in the shared program backbone.

This paper makes the following contributions:
\begin{itemize}
    \item We present an execution-driven structural abstraction for studying
    configuration-rich software under fuzzing, based on a shared function-level
    call graph and driver-induced subgraphs derived from dynamic execution.
    
    \item We develop lightweight tool support, {\mfuzz} and
    {\graphdist}, that operationalizes controlled multi-driver execution,
    strict per-driver runtime attribution, and offline graph-based structural
    analysis for empirical study.
    
    \item We develop a scalable empirical methodology and structural metric set
    for analyzing exploration scope, cohesion, fragmentation, modularity,
    overlap, and region-level under-exploration across execution modes.
    
    \item We conduct a large-scale empirical study on OSS-Fuzz-derived
    benchmarks and provide a systematic graph-level characterization of how
    multi-driver fuzzing explores software structure, including recurring
    residual-gap regimes and their implications for driver-aware fuzzing.
\end{itemize}

\vspace{3pt}
\noindent
\textbf{Open Science}
To support transparency, reproducibility, and follow-up research, we make the artifacts used in this study publicly available at \href{https://github.com/Cailbehumble/GraphDissect}{\underline{GraphDissect}}. These artifacts include: (i) the benchmark list and executable metadata, (ii) the manually specified driver configurations, and (iii) the scripts and tool support used for multi-driver fuzzing and structural analysis. 
We also document the experimental environment, execution workflow, and artifact organization to facilitate reuse by other researchers. By releasing both the execution-side and analysis-side artifacts, we aim to enable independent verification of our results and to support future work on driver-aware fuzzing, structural coverage analysis, and benchmark construction.
\section{Background \& Related Work}
\label{sec:background}

This section first introduces the background concepts underlying our study, 
including call graphs, fuzzing, and multi-mode execution, and 
then positions the study relative to prior work on fuzzing and program analysis.

\subsection{Program Structure and Call Graphs}
A call graph models the interprocedural structure of a program by representing functions as nodes and 
call relationships as directed edges~\cite{hall1992efficient}.
Call graphs are widely used to reason about program organization, modularity, and interprocedural behavior.
Depending on how they are constructed, 
call graphs may be static or dynamic, and context-sensitive or context-insensitive.
Static call graphs are typically derived through whole-program analysis and 
conservatively approximate the set of possible call relationships.
To preserve soundness, 
such analyses often over-approximate indirect calls caused by function pointers, callbacks, or virtual dispatch, especially in large C/C++ systems~\cite{hall1992efficient,murphy1998empirical,li2020pca}.
Dynamic call graphs, 
in contrast, record only the call edges exercised during execution, 
yielding precise but incomplete views of program behavior~\cite{graham1982gprof}.
In this work, 
call graphs provide the structural basis for reasoning about how different executions relate to the same program.
A function-level call graph offers a common granularity for comparing exercised program regions across heterogeneous software systems while remaining scalable to large real-world applications.

\subsection{Fuzzing and Dynamic Program Analysis}
Fuzzing is a widely used dynamic testing technique that 
generates large numbers of inputs to explore program behaviors and discover bugs~\cite{li2018fuzzing}.
Modern greybox fuzzers use lightweight instrumentation to 
guide input mutation through coverage feedback, 
such as basic-block or edge coverage~\cite{afl-fuzz,swiecki2016honggfuzz}.
This feedback-driven design has made fuzzing one of the most effective approaches 
for testing complex input-processing software.
At the same time, fuzzing observes program behavior only through concrete executions.
The coverage collected during fuzzing 
therefore reflects the portions of the program reached 
under a particular set of inputs and configurations~\cite{manes2019sok}, 
rather than the full set of statically possible behaviors.
As a result, 
fuzzing naturally yields partial but execution-grounded observations of program structure.
This execution-driven property is especially useful for studying 
how software is explored in practice.
Rather than attempting to reconstruct complete calling contexts or precise execution traces, 
one can use coverage collected during fuzzing as a scalable signal 
for which functional regions are actually exercised.

\subsection{CLI Programs and Multi-Mode Execution}
Many real-world software systems, especially command-line utilities, expose
multiple execution modes through subcommands, options, flags, configuration
settings, or input formats~\cite{schroder2022empirical}.
Such programs are also common among practical fuzzing targets, including many
projects in OSS-Fuzz and in our benchmark suite~\cite{ossfuzz}.
Examples include command-line tools such as \texttt{xpdf}, \texttt{jq},
\texttt{xmllint}, and \texttt{ffmpeg}, which provide distinct functionality
through different options, subcommands, or input modes~\cite{xpdf,jq,libxml2,ffmpeg}.

These invocation modes often activate different parsing logic, control-flow
paths, and downstream functionality within the same executable.
As a result, the code exercised under one mode may differ substantially from
that exercised under another.
This property is especially important in fuzzing, where program behavior depends
not only on mutated input bytes but also on how the target is invoked.
Compared with single-mode programs, CLI targets with rich mode-dependent
behavior therefore present a broader and more heterogeneous exploration space.
Understanding how these execution modes relate to exercised program structure is
important for analyzing overlap, diversity, and uneven exploration across
complex software systems.

\subsection{Related Work on Fuzzing and Program Analysis}
Prior work has explored the use of static and dynamic program analysis
to improve fuzzing effectiveness.
Existing techniques have used control-flow and data-flow information to identify
targets, prioritize exploration, guide mutations, or increase reachability to
difficult program locations~\cite{li2023polyfuzz}.
Other approaches combine fuzzing with symbolic execution or related hybrid
analyses to mitigate path explosion and improve behavioral coverage~\cite{zhang2020survey}.

Related work has also shown that the structure of a target program affects
fuzzing behavior.
Coverage feedback, control-flow organization, and input-dependent execution
diversity all influence which parts of a program are explored and which remain
difficult to reach~\cite{afl-fuzz,swiecki2016honggfuzz,zhu2022fuzzing}.
A closely related line of work studies fuzzing for command-line and
configuration-rich software, where program behavior depends not only on mutated
inputs but also on how the target is invoked.
Recent approaches have explored option-aware and CLI-oriented fuzzing, including
automatic extraction of program-option constraints from documentation, coverage-guided
combinatorial exploration of command-line interfaces, and interleaved fuzzing of
program options and file inputs~\cite{wang2023carpetfuzz,lu2024fuzzing,lee2025zigzagfuzz}.
More broadly, prior work in combinatorial interaction testing has shown the
importance of systematically exercising parameter combinations and configuration
constraints~\cite{yilmaz2012test,gargantini2016validation}.
These studies demonstrate that invocation modes and option combinations can
substantially affect coverage and bug discovery.
Related work has also considered fuzzing the broader program environment. Program environment fuzzing records and replays observed environmental interactions at the kernel/user-mode boundary, with selective mutations applied to explore different concrete program environments without explicit environment modeling ~\cite{Meng2024EnvFuzz}. This line of work shows that environmental interactions can affect program behavior and bug discovery. However, it primarily aims to extend fuzzing to richer environmental inputs for bug finding, rather than to characterize structural relationships among different execution modes of the same program.
Overall, these studies primarily treat options, execution modes, or environmental diversity as means to improve fuzzing effectiveness, rather than as objects of structural analysis.

Beyond fuzzing techniques, several empirical studies have examined fuzzing
behavior, benchmarking practice, and evaluation methodology.
Prior work has analyzed the reliability of fuzzing evaluations, the
reproducibility of coverage and bug-finding results, and the importance of
benchmark design in comparing fuzzers~\cite{klees2018evaluating,fuzzbench,magma,ossfuzz}.
These studies have improved understanding of how fuzzers should be evaluated,
but they mainly focus on aggregate outcomes such as coverage, crashes, and
benchmark quality, rather than on how different execution modes partition and
share the software structure of the same target.

Most prior studies therefore focus either on improving fuzzing guidance, extending fuzzing to richer inputs and environments, or on
evaluating fuzzers through aggregate effectiveness measures, rather than
explicitly analyzing the graph-level structural relationships among execution
modes of the same program.
Compared with these approaches, our work does not propose a new fuzzing
algorithm or guidance strategy.
Instead, it studies software structure as observed through fuzzing, using a
static call graph as a common structural backbone and dynamic execution
coverage as an execution-grounded signal of explored functionality.
This design emphasizes structural characterization rather than fuzzing
optimization.

\subsection{Positioning of This Study}
This study differs from prior work in both goal and methodology.
Rather than improving fuzzing guidance directly, we aim to understand how software structure is explored across different execution modes of the same program. 
Our focus is on how exercised program regions diverge, overlap, and remain under-explored under multi-mode fuzzing.
Methodologically, the study combines a unified static structural representation with execution-derived observations from fuzzing.
This combination makes it possible to compare exercised regions across drivers, executables, and benchmarks in a common graph space while keeping the analysis grounded in behavior that actually occurs during testing.
The resulting perspective complements existing work on fuzzing and program analysis by revealing structural properties that are difficult to observe from aggregate coverage alone.
\section{Study Goals and Overview}
\label{sec:goal}

The goal of this work is to empirically characterize how software structure is
exercised under different execution modes.
Modern software systems often expose rich functionality through command-line
options, configuration flags, and input formats.
Each configuration may activate a different subset of program behavior, yet the
structural relationships among these execution modes remain poorly understood.
This paper presents an empirical study of software structure as observed
through fuzzing.
Rather than proposing a new fuzzing or analysis technique, we focus on
measuring and analyzing the structural properties of real-world programs under
multi-driver execution.
We model software structure at the function level and study how different
drivers activate overlapping or distinct regions of the same program.
Our objective is to provide a systematic understanding of structural diversity,
sharing, and under-exploration in multi-driver software systems.

To enable consistent analysis across heterogeneous programs, we adopt a unified
graph-based representation.
For each target program, we construct a static, context-insensitive call graph
as a shared structural backbone.
Using fuzzing-driven dynamic profiling, we then identify, for each driver, the
subset of functions exercised during execution.
This yields a driver-induced subgraph for each execution mode, allowing direct
comparison of drivers using a common set of structural metrics.
Using this framework, we investigate the following research questions:

\begin{itemize}

\item \textbf{RQ1: Does multi-driver fuzzing achieve greater overall coverage than the best single driver under the same total fuzzing budget?}
We compare the union multi-driver results with those of the best individual
driver while keeping the total fuzzing budget constant.
Specifically, we evaluate both function-level structural coverage on the
call-graph backbone and final CFG edge coverage.

\item \textbf{RQ2: Do drivers exhibit heterogeneous fuzzing effectiveness under equal time budgets?}
We allocate the same fuzzing time to each driver and compare their individual
outcomes, including call-graph node coverage, CFG edge coverage, and discovered
bugs.

\item \textbf{RQ3: How are driver-induced subgraphs structurally organized?}
We analyze the size, connectivity, fragmentation, and modularity of
driver-induced subgraphs to examine how different execution modes activate and
organize program structure.

\item \textbf{RQ4: To what extent do drivers overlap in the functionality they cover?}
We measure pairwise overlap between drivers using intersection-over-union
(IoU) over covered functions and call edges.

\item \textbf{RQ5: Which structural regions remain under-explored after round-robin multi-driver fuzzing, and why?}
We identify residual low-coverage regions in the shared structural backbone
after multi-driver fuzzing and analyze the main causes of these gaps through
representative case studies.

\end{itemize}

These questions are designed to address the study goal from
complementary empirical perspectives.
Measuring overall effectiveness is necessary to determine whether multi-driver
execution changes exploration outcomes in a meaningful way.
However, aggregate outcomes alone are insufficient, because they do not reveal
whether the observed behavior is shared across drivers or dominated by only a
few execution modes.
For this reason, the study further examines variation across drivers and then
analyzes the structure and overlap of their induced subgraphs, so that
differences in exploration can be interpreted in terms of shared and distinct
program regions.
The analysis finally extends to residual under-explored structure, which is
necessary for understanding what parts of the shared backbone remain weakly
exercised after combining drivers.
In this way, the questions collectively connect outcome-level measurement with
structural characterization and residual-gap analysis to support the overall
empirical goal of understanding how software structure is exercised under
multi-driver fuzzing.

\section{Methodology} \label{sec:method}

This section presents the methodology of our empirical study.
We first introduce the overall study design and 
the structural abstraction used to analyze software structure under fuzzing.
We then describe the four study phases: benchmark selection, structural backbone construction, 
fuzzing and execution profiling, and structural analysis and evaluation.
Finally, we summarize the experimental setup and tool support used to operationalize the study.

\subsection{Study Design Overview}
\label{sec:studymethod}

This study empirically examines how software structure is exercised and 
explored under different execution modes.
We focus on understanding how distinct drivers activate overlapping or 
disjoint functional regions of a program, 
how these regions evolve during fuzzing, and 
where existing fuzzing campaigns leave structural gaps.
To this end, 
we model software structure at the function level and 
analyze driver-induced subgraphs derived from dynamic execution.

\begin{figure}[htp]
  \centering
  \includegraphics[width=1\textwidth]{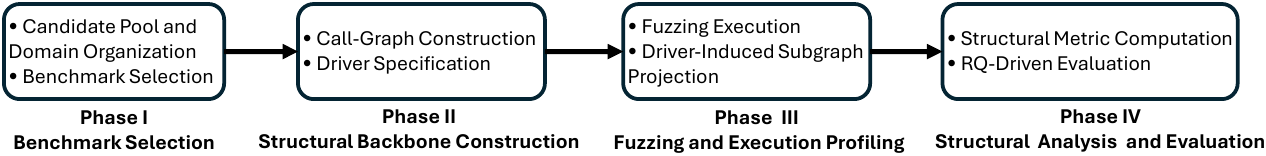}
  \caption{Overview of the empirical study methodology.}
  \label{fig:overview}
\end{figure}

Figure~\ref{fig:overview} summarizes the overall workflow of the study. 
The methodology proceeds in four phases. 
Phase I defines the benchmark space by organizing the candidate pool and selecting the final benchmarks. 
Phase II constructs a whole-program function-level call graph for each executable and 
specifies driver configurations representing distinct execution modes. 
Phase III runs fuzzing under each driver, collects driver-attributed execution profiles, and 
projects covered functions onto the shared call-graph backbone to obtain driver-induced subgraphs. 
Phase IV computes structural metrics over these subgraphs and organizes the resulting measurements to answer RQ1–RQ5.
This phased design enables systematic measurement of software structure from an execution-driven perspective.
By separating structural preparation, dynamic data collection, structural analysis, and evaluation, the study ensures that observations are both precise and comparable across programs.
Using a unified structural backbone allows driver-specific behaviors to be analyzed within a consistent representation, while dynamic profiling grounds the analysis in behaviors actually observed during fuzzing.
Together, these phases provide a scalable and principled framework for empirically studying software structure and fuzzing effectiveness across diverse real-world systems.

\subsection{Execution-Driven Structural Abstraction}\label{ssec:driver}

Rather than analyzing program structure solely as a static property, we focus on the
structural regions that are actually exercised during fuzzing.
This design choice is central to our methodology.
While static call graphs provide a program-wide view of potential interprocedural relationships,
they inevitably over-approximate runtime behavior, especially in large C/C++ systems
with indirect calls, callbacks, and configuration-dependent execution paths~\cite{hall1992efficient,murphy1998empirical,li2020pca}.
As a result, purely static structures often contain regions 
that are unreachable, infeasible, or irrelevant to the executions observed in practice.

Fuzzing, by contrast, explores programs through concrete executions generated under specific
inputs and execution modes.
From this perspective, 
the structure most relevant to fuzzing is not the full set of
statically possible relationships, 
but the subset that is actually activated during execution~\cite{manes2019sok}.
Grounding the study in execution-derived structure 
therefore allows us to analyze software
as fuzzing encounters it in practice: 
\textit{which regions are exercised, which execution modes
reach distinct functionality, where structural overlap exists across drivers,
and which regions remain persistently unexplored.}
Based on this view, we adopt a hybrid structural abstraction.
Each target program is represented by a function-level static call graph,
in which nodes denote functions and edges denote potential interprocedural call relationships.
This graph serves as a unified structural backbone shared across all drivers and fuzzing runs.
Dynamic execution profiles collected during fuzzing are then projected onto this backbone,
so that the study can compare structurally different executions within a common reference space.
This abstraction captures the interprocedural structure relevant to our analysis
while remaining lightweight enough for systematic comparison across benchmarks.

\begin{figure}[htp]
  \centering
  \includegraphics[width=0.75\textwidth]{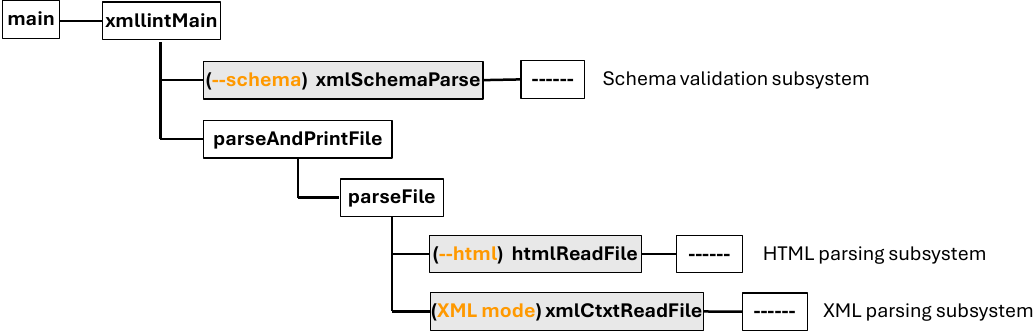}
  \caption{Example of driver-induced structural divergence in \texttt{xmllint}.
  All execution modes share a common structural backbone rooted at \texttt{xmllintMain}.
  Depending on the driver configuration, execution diverges into distinct functional regions,
  such as schema parsing (\texttt{--schema}), HTML parsing (\texttt{--html}), or XML parsing (default mode).
  Each driver thus induces a subgraph of the unified call graph, resulting in overlapping but
  non-identical structural coverage.}
  \label{fig:structuralExample}
\end{figure}

Figure~\ref{fig:structuralExample} illustrates this idea using \texttt{xmllint}~\cite{libxml2}.
Although the program has a single static structural backbone rooted at \texttt{xmllintMain},
different execution modes activate different functional regions.
For example, the default mode primarily exercises XML parsing logic,
whereas \texttt{--html} activates HTML parsing routines and \texttt{--schema}
further triggers schema-processing functionality.
Thus, even though all executions are embedded in the same program structure,
their effective exploration footprints differ substantially.

To represent these execution modes explicitly, we treat \emph{drivers} as the primary unit of analysis.
A driver corresponds to a concrete execution configuration of a target program,
defined by its options, expected input format, and seed corpus.
Formally, a driver configuration \( d \) is defined as
\[
d = \langle \mathit{id}, \mathit{opts}, \mathit{fmt}, S_d \rangle,
\]
where \( \mathit{id} \) is a unique driver identifier,
\( \mathit{opts} \) is the set of execution options and flags,
\( \mathit{fmt} \) is the expected input format,
and \( S_d \) is the seed corpus used to initialize fuzzing for that driver.
Each driver is intended to represent a semantically distinct execution mode 
and typically activates a characteristic subset of program functionality.
Under this abstraction, 
dynamic execution profiles collected during fuzzing are attributed
to driver identifiers and later mapped onto the shared call-graph backbone.
This yields \emph{driver-induced subgraphs}, 
which capture the effective structural regions explored by each driver.
These subgraphs form the basis of the subsequent analysis:
they allow us to quantify structural diversity across drivers,
measure overlap and imbalance in explored regions,
and relate structural exploration patterns to fuzzing outcomes such as coverage and crashes.

Overall, this hybrid design combines the strengths of static and dynamic views.
The static call graph provides a consistent program-wide reference,
while dynamic execution profiles ensure that the analysis remains grounded in
behavior that fuzzing actually exercises.
This combination enables a systematic empirical study of how software structure
is explored, shared, and missed across different drivers and benchmarks.

\subsection{Phase I: Benchmark Selection}\label{sec:phase1}

This phase defines the benchmark space and study scope.
Its goal is to select a benchmark set that is broad enough to capture
structural variation across software domains while remaining focused on
programs whose execution modes can be meaningfully represented as drivers.

\subsubsection{Candidate pool and domain organization}

We base our empirical study on real-world open-source software drawn from OSS-Fuzz~\cite{ossfuzz},
a large-scale continuous fuzzing platform maintained by Google.
OSS-Fuzz provides standardized build environments, fuzzing infrastructure, and security-relevant target programs,
making it a suitable and widely used source for empirical fuzzing studies.
From the full OSS-Fuzz project pool, which contains 1325 projects at the time of our study,
we restrict the benchmark space to programs implemented primarily in C or C++.
This choice is motivated by two considerations.
First, C/C++ projects are a natural focus within OSS-Fuzz, where C/C++ is directly supported and analysis support is especially mature~\cite{ossfuzz}.
Second, our study requires reliable whole-program analysis and function-level instrumentation, both of which are well supported in the LLVM/Clang ecosystem for C/C++~\cite{lattner2004llvm,sui2016svf}.
The resulting C/C++ project set contains 501 projects and 
forms the candidate pool for subsequent domain analysis and benchmark selection.

\begin{table}[htp]
\centering
\small
\caption{Domain distribution of C/C++ OSS-Fuzz~\cite{ossfuzz} projects used for candidate benchmark selection.
Projects are manually grouped by dominant application domain.
Code size is measured in KLoC (thousand lines of code) and reported as the observed min--max range per domain (excluding zero-LOC entries).}
\label{tab:domains}
\begin{tabular}{r l r r p{5.0cm}}
\toprule
\textbf{No.} & \textbf{Domain} & \textbf{\#Projects} & \textbf{Code Size Range (KLoC)} & \textbf{Domain Description} \\
\midrule

1  & Network and protocols
   & 170 & 2.44--7,942.90
   & Network-facing software processing protocols and remote inputs. \\

2  & Media processing
   & 94  & 2.62--2,241.49
   & Audio, video, and image codecs and processing pipelines. \\

3  & Metadata and system utilities
   & 58  & 4.69--1,934.10
   & Libraries and utilities for configuration, logging, and system metadata. \\

4  & Parsing and document processing
   & 69  & 2.96--5,114.92
   & Grammar- and format-driven parsers for structured data and documents. \\

5  & Toolchain / binary utilities
   & 22  & 7.16--9,542.76
   & Compilers, linkers, and tools operating on binaries or intermediate representations. \\

6  & Archive and compression
   & 22  & 4.63--274.44
   & Compression libraries and archive manipulation utilities. \\

7  & Language runtimes / interpreters
   & 23  & 2.60--4,113.27
   & Language runtimes and interpreters executing user programs. \\

8  & Database and storage
   & 16  & 7.63--6,064.44
   & Storage engines and libraries for persistent data management. \\

9  & OS / system software
   & 20  & 3.82--1,709.08
   & Operating-system components and low-level system services. \\

10 & Machine learning
   & 7   & 52.33--3,171.45
   & Machine-learning frameworks and numerical computation libraries. \\

\midrule
\textbf{Total} &  & \textbf{501} & \textbf{2.44--9,542.76} & -- \\
\bottomrule
\end{tabular}
\end{table}

To ensure diversity in application characteristics, 
we manually organize the candidate projects by dominant application domain.
This grouping provides a coarse-grained but useful view of the candidate pool:
in empirical software engineering, 
the diversity and type of studied systems affect how broadly findings can be interpreted, 
and software taxonomies help
relate evidence to its context~\cite{nagappan2013diversity,forward2008taxonomy,baltes2022sampling}.
The domain categories used here therefore emphasize dominant functionality and
input-processing role, which are the aspects most relevant to driver-based
structural exploration under fuzzing.
In addition, prior fuzzing benchmark efforts have emphasized the importance of
using diverse, representative, and real-world targets to avoid overfitting and
to improve the realism of evaluation~\cite{fuzzbench,magma}.
Each project is assigned to exactly one domain according to its primary
artifact, dominant entry-point behavior, and core functionality.
To improve labeling consistency, the assignment is based on project
documentation, repository descriptions, and the purpose of the fuzzed
executable, with ambiguous cases revisited through joint inspection of these
sources.
When a project spans multiple functionalities, it is assigned to the domain
that best reflects its dominant use case and execution behavior.
Table~\ref{tab:domains} summarizes the resulting domain distribution.

\subsubsection{Benchmark Selection}

\begin{table}[htp]
\centering
\small
\caption{Benchmarks used in the empirical evaluation.
All programs are selected from OSS-Fuzz and expose multiple execution modes
through command-line options or input formats.}
\label{tab:benchmarks}
\begin{tabular}{l l r l r r}
\toprule
\textbf{Domain} & \textbf{Benchmark} & \textbf{LOC (K)} & \textbf{Executable} & \textbf{\#Functions} & \textbf{\#Drivers} \\
\midrule

\multirow{5}{*}{Network and protocols}
& \multirow{2}{*}{snort3~\cite{snort3}} & \multirow{2}{*}{367.1}
    & snort       & 50,881 & 9 \\
&  & 
    & snort2lua   & 4,528  & 10 \\
& unbound~\cite{unbound} & 137.6
    & checkconf   & 2,786  & 7 \\
& \multirow{2}{*}{http-parser~\cite{http_parser}} & \multirow{2}{*}{6.2}
    & parsertrace & 26     & 3 \\
&  &
    & url\_parser & 20     & 2 \\\hline

\multirow{8}{*}{Media processing}
& \multirow{2}{*}{ffmpeg~\cite{ffmpeg}} & \multirow{2}{*}{1,348.5}
    & ffmpeg     & 29,709 & 19 \\
&  &
    & ffprobe    & 29,303 & 28 \\
& \multirow{3}{*}{libtiff~\cite{libtiff}} & \multirow{3}{*}{91.9}
    & tiff2bw    & 679    & 7 \\
&  &
    & tiffinfo   & 697    & 7 \\
&  &
    & tiff2pdf   & 818    & 21 \\
& \multirow{3}{*}{wavpack~\cite{wavpack}} & \multirow{3}{*}{85.6}
    & wavpack    & 310    & 32 \\
&  &
    & wvunpack   & 206    & 30 \\
&  &
    & wvgain     & 184    & 12 \\\hline

\multirow{5}{*}{Metadata and system utilities}
& git~\cite{git} & 307.0
    & git         & 12,714 & 6 \\
& \multirow{3}{*}{sleuthkit~\cite{sleuthkit}} & \multirow{3}{*}{284.6}
    & istat       & 5,027  & 4 \\
&  &
    & img\_stat   & 393    & 7 \\
&  &
    & tsk\_recover & 5,367 & 4 \\
& file~\cite{file_cmd} & 188.0
    & file        & 331    & 12 \\\hline

\multirow{5}{*}{Parsing and document processing}
& \multirow{3}{*}{xpdf~\cite{xpdf}} & \multirow{3}{*}{493.2}
    & pdfdetach & 2,686 & 12 \\
&  &
    & pdfinfo   & 2,712 & 25 \\
&  &
    & pdftops   & 3,001 & 66 \\
& libxml2~\cite{libxml2} & 139.1
    & xmllint   & 2,796 & 40 \\
& jq~\cite{jq} & 32.4
    & jq        & 800   & 18 \\\hline

\multirow{5}{*}{Toolchain / binary utilities}
& \multirow{3}{*}{binutils~\cite{binutils}} & \multirow{3}{*}{3,134.9}
    & objdump   & 3,993 & 22 \\
&  &
    & readelf   & 1,551 & 29 \\
&  &
    & addr2line & 2,825 & 8 \\
& cppcheck~\cite{cppcheck} & 278.4
    & cppcheck  & 37,281 & 48 \\
& libdwarf~\cite{libdwarf} & 129.4
    & dwarfdump & 1,592 & 29 \\\hline

\multirow{4}{*}{Language runtimes / interpreters}
& cpython3~\cite{cpython} & 763.7
    & python & 13,953 & 23 \\
& \multirow{2}{*}{quickjs~\cite{quickjs}} & \multirow{2}{*}{76.3}
    & qjs  & 1,944 & 6 \\
&  &
    & qjsc & 1,947 & 27 \\
& lua~\cite{lua} & 21.3
    & lua  & 1,148 & 36 \\\hline

\multirow{4}{*}{Archive and compression}
& \multirow{2}{*}{libarchive~\cite{libarchive}} & \multirow{2}{*}{168.9}
    & bsdtar   & 2,223 & 15 \\
&  &
    & bsdunzip & 606   & 13 \\
& upx~\cite{upx} & 144.3
    & upx      & 5,824 & 22 \\
& xz~\cite{xz} & 34.4
    & xz       & 749   & 17 \\\hline

\multirow{7}{*}{Database and storage}
& \multirow{3}{*}{hdf5~\cite{hdf5}} & \multirow{3}{*}{739.4}
    & h5dump   & 5,643 & 12 \\
&  &
    & h5ls     & 5,624 & 12 \\
&  &
    & h5repack & 5,632 & 37 \\
& \multirow{3}{*}{netcdf~\cite{netcdf}} & \multirow{3}{*}{296.6}
    & ncdump  & 2,476 & 17 \\
&  &
    & ncgen   & 2,721 & 22 \\
&  &
    & nccopy  & 2,404 & 30 \\
& sqlite3~\cite{sqlite3} & 289.5
    & sqlite3 & 2,636 & 48 \\

\midrule
\textbf{Total} & 27 & 9,262.0 & 43 & 258,746 & 854 \\
\bottomrule
\end{tabular}
\end{table}

From the domain-organized candidate pool, 
we select a subset of benchmarks for detailed study.
The selection is guided by three criteria.
First, we seek \emph{scale diversity} by choosing projects that span
different code sizes within each domain.
Second, we require \emph{execution-mode richness}, so that each selected
target exposes multiple meaningful execution modes through command-line
options or subcommands, and therefore supports driver-based analysis.
Third, we prioritize \emph{practical relevance}, 
favoring actively
maintained and widely used projects whose structure and security exposure
reflect realistic fuzzing targets.
This strategy yields a benchmark set that balances breadth and depth.
It preserves diversity across domains, architectural styles, 
and program scales while keeping the set manageable for detailed structural analysis.
The final benchmark set is shown in Table~\ref{tab:benchmarks}.

The selected benchmarks reflect a broad class of user-space
input-processing software commonly fuzzed in practice.
Across domains, 
they exhibit the characteristics most relevant to this study: 
multiple execution modes, non-trivial internal structure, option-driven behavior, 
and sufficiently rich functional decomposition to induce distinct structural footprints.
As a result, 
the structural phenomena examined here, 
including driver-induced divergence, overlap, and coverage imbalance, 
are observed across diverse software systems rather 
than being tied to a single application class.
We intentionally exclude categories such as operating-system kernels,
low-level firmware, and machine-learning frameworks.
These systems are less aligned with the central abstraction of this study,
which focuses on execution-mode-dependent structural exploration through
explicit driver configurations.
By contrast, user-space applications and utilities more naturally expose
interchangeable execution modes through options, formats, and subcommands,
making them a better fit for analyzing how fuzzing interacts with software
structure across drivers.
\subsection{Phase II: Structural Backbone Construction}\label{sec:phase2}

In Phase II, we prepare a unified structural representation for each target executable and define the driver configurations used throughout the study.

\subsubsection{Call-Graph Construction}
This component constructs a whole-program, function-level call graph
that serves as the unified structural backbone for analysis.
We obtain call graphs through static analysis of the target program,
resolving both direct and indirect call relationships using the
flow-sensitive pointer analysis component of SVF~\cite{sui2016sparse}.
This analysis conservatively approximates possible call targets,
yielding a call graph that captures potential interprocedural
relationships among functions without relying on execution-specific information.

A static call graph provides a stable and consistent reference structure
that is shared across all drivers and fuzzing runs.
Although such graphs may over-approximate feasible execution paths,
they enable systematic comparison of driver-induced subgraphs across
execution modes and benchmarks.
Moreover, this design avoids the substantial instrumentation,
runtime overhead, and scalability limitations associated with
context-sensitive or trace-level representations,
making it suitable for large-scale empirical analysis.
Importantly, 
our study does not require exact call precision, 
as all structural measurements are grounded 
in dynamically observed coverage projected onto this static backbone.

\subsubsection{Driver Specification}
Based on the driver abstraction defined in Section~\ref{ssec:driver}, we manually specify
the concrete driver configurations used in the study.
Each driver is instantiated as a fixed execution configuration with a unique identifier,
together with its command-line options, expected input format, and seed corpus.
For each target program, we maintain a one-to-one mapping between driver identifiers
and their corresponding execution parameters.
During fuzzing, each execution is launched under one such configuration,
and all produced artifacts, including coverage and crashes, are attributed to the corresponding driver identifier.
This operationalization ensures that later profiling and structural analysis remain driver-specific throughout the pipeline.

Manual specification ensures that each driver corresponds to a semantically meaningful execution mode
rather than a superficial variation of inputs.
It also avoids ambiguities that may arise from automated option inference or dynamically changing configurations,
thereby improving reproducibility, interpretability, and reliable cross-driver comparison.
In addition, each driver configuration is manually validated before inclusion to ensure that it is executable under the target program,
accepts the intended input form, and can serve as a stable basis for fuzzing.
We intentionally focus driver construction on primary command-line options and major CLI modes
rather than exhaustively enumerating all possible option combinations or other execution-shaping factors.
This restriction keeps the driver space interpretable and comparable across heterogeneous programs,
while allowing the study to focus on execution modes that are more likely to correspond to distinct functional behavior.
The resulting drivers therefore represent a controlled and representative abstraction for empirical analysis.

\begin{figure}[htp]
  \centering
  \includegraphics[width=0.7\textwidth]{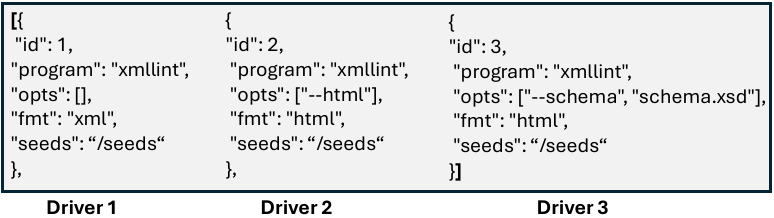}
  \caption{Example JSON specifications for three \texttt{xmllint} driver configurations
corresponding to XML parsing, HTML parsing, and XML schema validation.
Each driver is assigned a unique identifier and a fixed seed corpus.}
  \label{fig:driverjson}
\end{figure}

\vspace{3pt}
\noindent
\textbf{Example.}
We illustrate driver specification using \texttt{xmllint} in Figure~\ref{fig:driverjson}, 
where different command-line modes
activate distinct parsing and validation components (Figure~\ref{fig:structuralExample}).
In our study, each driver is recorded as a JSON object
\(\langle \mathit{id}, \mathit{opts}, \mathit{fmt}, S_d \rangle\),
where \texttt{id} uniquely identifies the execution mode,
\texttt{opts} encodes the command-line configuration,
\texttt{fmt} denotes the intended input format,
and \texttt{seeds} points to a fixed seed corpus used to initialize fuzzing.
The first driver targets the default XML parsing mode.
The second driver enables HTML parsing via \texttt{--html}, inducing a distinct structural region.
The third driver enables schema validation by providing \texttt{--schema} together with a schema file,
which triggers schema compilation and validation routines in addition to XML parsing.
In all cases, the fuzzer starts from the corresponding seed corpus and attributes all observed
coverage and crashes to the driver identifier.

\subsection{Phase III: Fuzzing and Execution Profiling}\label{sec:phase3}

In Phase III, 
we run fuzzing under the configured drivers and 
collect the
execution profiles used in later structural analysis.
The goal of this phase is to expose each target program 
to multiple execution modes under a controlled 
and comparable setting 
while recording the driver-specific runtime artifacts needed for subsequent analysis.

\subsubsection{Fuzzing Execution}

For each target program, 
we conduct fuzzing across the configured set of
drivers under a fixed shared time budget.
Each driver is treated as a distinct execution mode, 
and the total fuzzing time is divided equally across all drivers so that each one is evaluated under comparable resource constraints.
Driver execution follows a round-robin policy, 
where fuzzing iterates through the driver set and activates one configuration at a time.
When a driver is selected, 
fuzzing proceeds under its associated execution parameters, 
including command-line options and seed corpus.
During its assigned time slice, 
the fuzzer repeatedly generates inputs and executes the target program under that configuration.
This process is repeated across all drivers until the total fuzzing budget is exhausted.
By using the same fuzzing engine, mutation workflow, and execution environment throughout the study,
we ensure that differences observed across drivers are attributable to the execution modes themselves rather than to variations in experimental setup.

This execution policy supports fair cross-driver comparison in two ways.
First, equal time allocation prevents the observed results from being dominated by a small subset of drivers that receive more resources.
Second, round-robin scheduling ensures that all drivers are exercised under the same overall fuzzing conditions,
which improves the comparability of their coverage, crash outcomes, and structural footprints.

\subsubsection{Driver-Induced Subgraph Projection}

While fuzzing is in progress, 
we continuously collect driver-specific execution profiles for later analysis.
These profiles include the runtime artifacts needed to relate fuzzing behavior to software structure,
including covered functions, covered control-flow edges, and discovered crashes.
All such artifacts are attributed to the active driver configuration under which the execution occurs.
To preserve this attribution, 
profiling is maintained separately for each driver throughout the fuzzing process.
Coverage and crash artifacts observed during a driver's execution slice are
recorded under that driver identifier.
The covered functions are then projected to the shared whole-program call graph introduced in Phase II (Section~\ref{sec:phase2}),
yielding a driver-specific execution footprint that can be analyzed in later phases.
In this way, Phase II converts fuzzing executions into driver-attributed structural observations.

This profiling design is important for the goals of the study.
Because the analysis focuses on how different execution modes explore overlapping or distinct program regions,
runtime artifacts must remain attributable to individual drivers rather than aggregated across the program as a whole.
Maintaining per-driver profiles therefore enables later phases to compare structural diversity, overlap, imbalance,
and uncovered regions across execution modes in a principled way.
\subsection{Phase IV: Structural  Analysis  and Evaluation
}\label{ssec:phase4}

Phase IV quantifies how fuzzing explores software structure under different drivers.
Given the driver-induced subgraphs obtained in Phase III (Section~\ref{sec:phase3}), 
this phase measures their size, internal organization, overlap, and region-level exploration patterns.
The goal is not only to describe the structural footprints reached by individual drivers,
but also to understand how these footprints differ across execution modes and how they
relate to fuzzing outcomes.
All measurements are computed on \emph{driver-induced subgraphs} derived from observed executions,
using the whole-program static call graph as a shared structural reference.
This design enables comparisons across drivers and executables in a common graph space,
while keeping the analysis grounded in structure that fuzzing actually exercises.

\subsubsection{Structural Metric Computation}\label{ssec:metrics}

Let the whole-program static call graph be $G=(V,E)$.
For a given driver $d$, let its driver-induced subgraph be
$G_d=(V_d,E_d)$, where $V_d \subseteq V$ and $E_d \subseteq E$
denote the functions and call edges activated by fuzzing through driver $d$.
To analyze structural exploration systematically, we organize the metrics into five purposes:
(1) measuring exploration scope,
(2) characterizing cohesion and fragmentation,
(3) capturing internal community organization,
(4) quantifying redundancy and complementarity across drivers, and
(5) identifying under-explored structural regions.

\vspace{3pt}
\noindent
\textit{\blackcircleone{1} Metrics for exploration scope.}
To measure how much program structure a driver reaches, we use subgraph size.

\noindent
\textbf{\textit{Subgraph size.}}
For each driver-induced subgraph $G_d$, we measure its size using the number of activated functions and call edges,
i.e., $|V_d|$ and $|E_d|$.
This metric captures the structural footprint of a driver.
A larger value indicates that the driver reaches a broader interprocedural region of the program,
whereas a smaller value indicates that execution remains confined to a narrower structural footprint.
We include this metric because exploration size is the most basic indicator of how much structure has been exercised,
and it provides a natural baseline for interpreting the more detailed structural measurements below.

\vspace{3pt}
\noindent
\textit{\blackcircleone{2} Metrics for structural cohesion and fragmentation.}
To distinguish whether a driver explores one dominant structural region or several scattered ones,
we measure both the dominance of the largest connected region and the total number of disconnected regions.

\noindent
\textbf{\textit{Largest Connected Component (LCC) ratio.}}
To measure the cohesion of a driver-induced subgraph, 
we compute the fraction
of activated functions contained in its largest weakly connected component~\cite{newman2003structure}:
\[
\mathrm{LCC}(G_d)=\frac{|V_{\mathrm{LCC}}|}{|V_d|},
\]
where $V_{\mathrm{LCC}}$ is the vertex set of the largest weakly connected component in $G_d$.
This metric indicates how strongly the explored structure is concentrated in one main connected region.
A larger LCC ratio suggests that most activated functions belong to a single dominant subsystem,
whereas a smaller ratio suggests that execution is distributed across multiple disconnected or weakly connected regions.

\noindent
\textbf{\textit{Number of Weakly Connected Components (\#WCC).}}
To measure fragmentation directly, we compute
\[
\#\mathrm{WCC}(G_d)=\text{number of weakly connected components in } G_d.
\]
This metric complements the LCC ratio.
Whereas LCC shows how dominant the largest region is, \#WCC shows how many distinct structural regions are activated in total.
A larger value indicates more fragmented exploration,
while a value of one indicates that all activated nodes belong to a single connected region.
Together, these two metrics provide a clearer picture of structural cohesion than either alone.

\vspace{3pt}
\noindent
\textit{\blackcircleone{3} Metric for internal community organization.}
To characterize whether the explored structure is internally organized into distinct functional clusters,
we measure modularity.

\noindent
\textbf{\textit{Modularity ($Q$).}}
Because the whole-program call graph and each driver-induced subgraph are directed,
we measure internal community organization using directed modularity.
Let $A_{ij}$ denote the adjacency matrix of $G_d$, where $A_{ij}=1$ if there is
a call edge from node $i$ to node $j$ and $A_{ij}=0$ otherwise.
Let $k_i^{\mathrm{out}}$ and $k_j^{\mathrm{in}}$ denote the out-degree of node $i$
and the in-degree of node $j$, respectively, and let $c_i$ denote the community
assignment of node $i$.
We compute directed modularity as:
\[
Q(G_d)=
\frac{1}{|E_d|}
\sum_{i,j}
\left(
A_{ij}-\frac{k_i^{\mathrm{out}}k_j^{\mathrm{in}}}{|E_d|}
\right)\delta(c_i,c_j),
\]
where $\delta(\cdot)$ is the Kronecker delta~\cite{leicht2008community}.
Community detection is performed using a Louvain-style heuristic that maximizes
directed modularity~\cite{hendrickx2008graphs,dugue2022direction}.
This metric captures an aspect of structural organization not reflected by size
or connectivity alone.
A larger modularity value indicates that the explored subgraph is organized
into more distinct directed communities with sparser cross-community links,
whereas a smaller value indicates weaker community structure.
We include this metric because two drivers may cover similarly large
subgraphs while differing substantially in how clearly their explored
structure is partitioned into directed functional clusters.

\vspace{3pt}
\noindent
\textit{\blackcircleone{4} Metrics for inter-driver redundancy and complementarity.}
To understand whether different drivers explore similar or distinct structural regions,
we measure pairwise overlap between their induced subgraphs.

\noindent
\textbf{\textit{Inter-driver structural overlap.}}
To quantify structural redundancy and complementarity across drivers, we compute pairwise intersection-over-union (IoU) between their induced subgraphs at both the function and call-edge levels:
\[
\mathrm{IoU}_{V}(d_i,d_j)=
\frac{|V_i \cap V_j|}{|V_i \cup V_j|}, \qquad
\mathrm{IoU}_{E}(d_i,d_j)=
\frac{|E_i \cap E_j|}{|E_i \cup E_j|}.
\]
Here, \(\mathrm{IoU}_{V}\) measures the overlap between the function sets of drivers \(d_i\) and \(d_j\),
while \(\mathrm{IoU}_{E}\) measures the overlap between their call-edge sets.
A larger IoU value indicates that two drivers activate more similar structural regions,
suggesting greater redundancy.
A smaller IoU value indicates that the two drivers cover more distinct regions,
suggesting greater complementarity.
Function-level IoU captures overlap in activated functions,
whereas edge-level IoU captures similarity in interprocedural execution structure.
These metrics are useful because the benefit of multi-driver fuzzing depends not only on
how much structure each driver covers individually, but also on how much new structure
it contributes relative to other drivers.

\vspace{3pt}
\noindent
\textit{\blackcircleone{5} Metrics for identifying under-explored structural regions.}
To support RQ5, we identify coarse-grained structural regions that remain
weakly explored after union multi-driver fuzzing.

\noindent
\textbf{\textit{Structural regions.}}
We partition the whole-program call graph into structural regions that serve
as coarse-grained analysis units:
\[
\mathcal{R}=\textsc{CommunityDetect}(G)=\{R_1,R_2,\dots,R_k\},
\]
where each region $R_i \subseteq V$ is a densely connected directed cluster.
In practice, we use a Louvain-style local-moving and aggregation heuristic
adapted to maximize directed modularity~\cite{blondel2008fast,dugue2015directed}.
These regions are used only as structural analysis units for locating
residual exploration gaps.

\noindent
\textbf{\textit{Region coverage.}}
Let $C_{\text{union}} \subseteq V$ denote the set of functions covered by
union multi-driver fuzzing.
For each structural region $R_i$, we define
\[
\mathrm{RC}(R_i)=\frac{|R_i \cap C_{\text{union}}|}{|R_i|}.
\]
This metric measures the fraction of functions in region $R_i$ that are
reached by union multi-driver fuzzing.
A smaller value indicates that the region remains less explored.
Regions with lower coverage are treated as candidate residual gaps for
later inspection.

Together, these region-level metrics reveal whether fuzzing explores the program broadly
or instead concentrates repeatedly on a limited structural core.
This perspective is important because aggregate coverage alone cannot show whether the observed progress
is evenly distributed across the program or confined to a small subset of regions.

\subsubsection{RQ-Driven Evaluation}

In Phase IV, we aggregate the fuzzing outcomes and structural measurements
obtained in the previous phases to answer the research questions.
The purpose of this phase is to connect the collected execution profiles and
driver-induced structural analyses to the empirical findings of the study.
To this end, we organize the evaluation according to the role of each
question.

RQ1 examines overall multi-driver effectiveness by comparing union
multi-driver results against individual-driver baselines using function
coverage, control-flow edge coverage, and crashes.
RQ2 evaluates whether fuzzing contributions are balanced across drivers by
analyzing per-driver outcome distributions and normalized contributions under
the shared execution policy.
RQ3 characterizes the structural properties of driver-induced subgraphs using
metrics such as size, connectedness, fragmentation, and modularity.
RQ4 measures structural redundancy and complementarity across drivers using
pairwise function-level and edge-level overlap.
RQ5 identifies structural regions that remain under-explored after union
multi-driver fuzzing using region coverage, and then uses representative case
studies to investigate the causes of these residual exploration gaps.

Organizing the evaluation in this way ensures that each research question is
answered by the measurements most directly relevant to it.
It also makes the connection between the study design and the reported
findings explicit, improving the interpretability of the overall empirical
analysis.

\subsection{Tool Support} \label{sec:tool}

\begin{figure}[htp]
  \centering
  \includegraphics[width=1\textwidth]{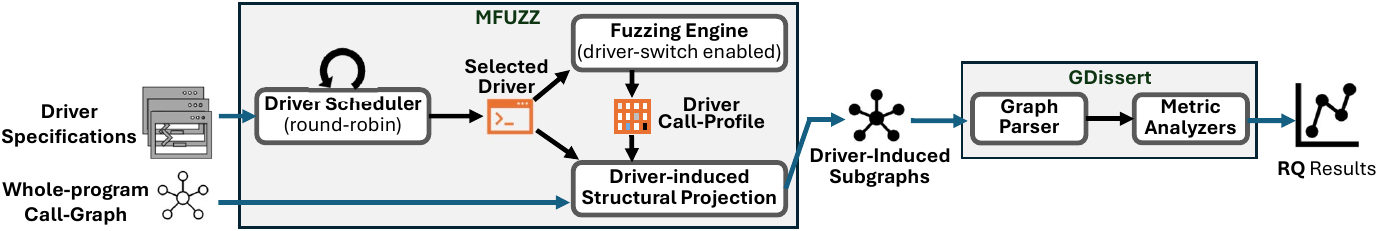}
  \caption{Tool support for the empirical study.
{\mfuzz} performs multi-driver fuzzing with driver switching and online
profile collection.
{\graphdist} performs offline analysis of driver-induced subgraphs and
computes the structural metrics used in the study.}
  \label{fig:tool}
  \vspace{-10pt}
\end{figure}

To implement the study pipeline, we develop two lightweight tool components,
{\mfuzz} and {\graphdist}, as shown in Figure~\ref{fig:tool}.
{\mfuzz} is the execution-side component. It extends Honggfuzz~\cite{swiecki2016honggfuzz}
with driver switching, shared-memory-based profile collection, and online
projection of per-driver execution onto the shared structural backbone.
{\graphdist} is the analysis-side component. It consumes the driver-induced
subgraphs produced by {\mfuzz} and performs offline parsing, aggregation, and
metric computation for the RQ-driven evaluation.
This separation is intentional.
{\mfuzz} is responsible for controlled multi-driver execution and runtime
artifact collection, whereas {\graphdist} performs post hoc structural
analysis.
As a result, the analysis remains observational and does not introduce
guidance or feedback into the fuzzing loop.
Additional implementation details are provided in the artifact.

\subsubsection{{\mfuzz}: Execution-side support}

{\mfuzz} is the execution-side implementation of the framework.
It is built on top of honggfuzz~\cite{swiecki2016honggfuzz}
and extends the base fuzzing engine with multi-driver scheduling
and online structural profiling.
Its purpose is to execute multiple driver configurations within a single
fuzzing instance while preserving strict attribution of runtime artifacts to
the active driver.

Given a total fuzzing budget $T$ and a driver set $\mathcal{D}$, {\mfuzz}
implements round-robin driver scheduling with equal time slices.
When a driver is selected, {\mfuzz} applies its execution configuration,
including command-line options and seed corpus, and switches the fuzzing
engine to that configuration until the current time slice ends.
To preserve driver attribution during execution, {\mfuzz} maintains
shared-memory coverage structures that are synchronized at driver-switch
boundaries.
During fuzzing, function- and call-edge coverage updates are recorded in the
shared profile together with the currently active driver identifier.
At the end of each time slice, {\mfuzz} snapshots the accumulated profile of
the outgoing driver and projects the observed coverage onto the shared
whole-program call graph.
This step yields the driver-induced subgraph used in later structural
analysis.
Internally, driver attribution is represented through driver-indexed bitmaps
attached to call-graph nodes and edges.
A set bit indicates that the corresponding function or call edge has been
observed under that driver in at least one execution during the current run.
This representation keeps profiling and projection lightweight, supports
efficient union and comparison across drivers, and avoids maintaining
separate full graph copies for each driver.

Using a unified fuzzing engine is an intentional implementation choice.
It ensures that all drivers share the same mutation logic, instrumentation,
scheduling framework, and runtime environment, reducing variance that would
otherwise arise from launching fully independent fuzzing instances.
At the same time, synchronized driver switching and shared-memory profiling
preserve strict per-driver attribution of coverage and crashes.
The implementation of {\mfuzz} is lightweight, consisting of approximately
4KLoC lines of C/C++ code on top of honggfuzz~\cite{swiecki2016honggfuzz}.

\subsubsection{{\graphdist}: Analysis-side support}

{\graphdist} is the analysis-side implementation of the framework.
It operates offline on the structural artifacts produced by {\mfuzz},
parses the resulting driver-induced subgraphs, and computes the measurements
used in the RQ-driven evaluation.
Implementation-wise, {\graphdist} is organized into two stages.
The first stage parses the whole-program call graph and the driver-induced
subgraph artifacts emitted by {\mfuzz}, reconstructing the per-driver
structural views needed for analysis.
The second stage applies the metric analyzers to compute the measurements
used throughout the study, including subgraph size, connectivity,
fragmentation, modularity, inter-driver overlap, and region coverage for
identifying under-explored structural regions.
These outputs support the quantitative analyses in RQ1--RQ5 and the
selection of representative cases for analyzing residual exploration gaps.
Because all measurements are computed with respect to the same whole-program
call graph, the resulting structural properties are directly comparable
across drivers and executables.
The implementation of {\graphdist} consists of approximately 2.8KLoC
lines of Python code, excluding third-party parsing and analysis libraries.

\subsubsection{Implementation Validation}

Because {\mfuzz} operates inside the fuzzing loop, we validate whether its
driver-switching and profiling mechanisms materially affect fuzzing behavior
or introduce substantial execution overhead.
To make this validation both representative and conservative, we select
\emph{one large executable per benchmark domain} from
Table~\ref{tab:benchmarks}, prioritizing targets with high function counts.
This choice stresses the main cost drivers of {\mfuzz}, since larger targets
generally require more structural state to maintain and project during
switching.
At the same time, selecting one target from each domain preserves diversity
across program classes, input formats, and execution structures.
The resulting validation set consists of \texttt{snort}, \texttt{ffprobe},
\texttt{git}, \texttt{pdftops}, \texttt{cppcheck}, \texttt{python},
\texttt{upx}, and \texttt{h5dump}.
Unless otherwise noted, all validation results are reported as averages over
five independent runs.

\vspace{3pt}
\noindent
\textbf{Coverage preservation under switching.}
To assess whether driver switching materially perturbs fuzzing outcomes, we
compare pure single-driver fuzzing with honggfuzz~\cite{swiecki2016honggfuzz}
against round-robin execution in {\mfuzz} over duplicated copies of the same
driver configuration.
For each target, we select the driver configuration that achieves the highest
single-driver CFG-edge coverage in the single-driver baseline used for RQ1,
and duplicate that same configuration in {\mfuzz}.
This avoids arbitrary option selection, aligns the validation with the
paper’s primary baseline, and ensures that the tested configuration
corresponds to a semantically meaningful execution mode already used in the
benchmark suite.
Since the duplicated drivers are semantically identical, any substantial
difference in final coverage can be primarily attributed to the switching
mechanism rather than to differences in execution mode.
We report the final number of covered CFG edges under the same fuzzing budget.
For readability, Table~\ref{tab:cov-preserve} reports only the
distinguishing driver options; the full commands, inputs, seed corpora, and
auxiliary files are provided in the artifact.
Small differences indicate that driver switching preserves the underlying
coverage behavior of the fuzzing process.

\begin{table}[htp]
\centering
\small
\caption{CFG-edge coverage under driver switching for the validation targets.
Reported values are averages over five independent runs. Relative difference
is computed as $(\text{{\mfuzz}}-\text{honggfuzz})/\text{honggfuzz}\times100$.}
\label{tab:cov-preserve}
\begin{tabular}{l p{2.5cm} r r r}
\toprule
\textbf{Executable} &
\textbf{Selected Option} &
\textbf{CFG Edges ({\mfuzz})} &
\textbf{CFG Edges (honggfuzz)} &
\textbf{Rel. Diff. (\%)} \\
\midrule
\texttt{snort}    & -c conf/min.lua         & 7407  & 7464  & -0.8\% \\
\texttt{ffprobe}  & -show\_streams -i       & 13070 & 12915 & 1.2\%  \\
\texttt{git}      & apply                   & 533   & 536   & -0.6\% \\
\texttt{pdftops}  & -level1                 & 2186  & 2228  & -1.9\% \\
\texttt{cppcheck} & \verb|--|enable=warning & 4113  & 4098  & 0.4\%  \\
\texttt{python}   & -x                      & 18709 & 18604 & 0.6\%  \\
\texttt{upx}      & -q compress             & 4958  & 4973  & -0.3\% \\
\texttt{h5dump}   & -n                      & 5388  & 5407  & -0.4\% \\
\bottomrule
\end{tabular}
\end{table}

Table~\ref{tab:cov-preserve} shows that single-driver honggfuzz and
round-robin execution in {\mfuzz} produce highly similar coverage results on
all eight validation targets.
The relative difference ranges from $-1.9\%$ to $+1.2\%$, and remains within
$1\%$ for six of the eight targets.
Moreover, the differences are mixed in sign rather than consistently favoring
either system, suggesting that they mainly reflect normal fuzzing variation
rather than a systematic effect of driver switching.
Overall, these results indicate that {\mfuzz} preserves coverage behavior
under switching with only minor deviation from pure single-driver fuzzing.

\vspace{3pt}
\noindent
\textbf{Execution overhead.}
We next quantify the practical overhead introduced by {\mfuzz} relative to
base honggfuzz on the same validation targets.
We report two measurements.
First, we measure the average cost of a driver switch, including
synchronization, profile snapshotting, and per-switch projection onto the
whole-program call graph.
Second, we report peak memory usage for both honggfuzz and {\mfuzz}, together
with their relative difference.

\begin{table}[htp]
\centering
\small
\caption{Execution overhead of {\mfuzz} relative to honggfuzz on the
validation targets. Reported values are averages over five independent runs.
Relative memory difference is computed as
$(\text{{\mfuzz}}-\text{honggfuzz})/\text{honggfuzz}\times100$.}
\label{tab:mfuzz-overhead}
\begin{tabular}{l r r r r}
\toprule
\textbf{Executable} &
\textbf{Avg. Switch Cost (ms)} &
\textbf{Peak Mem. ({\mfuzz}, MB)} &
\textbf{Peak Mem. (honggfuzz, MB)} &
\textbf{Rel. Diff. (\%)} \\
\midrule
\texttt{snort}    & 79.2  & 1019.7  & 833.2   & 22.4\% \\
\texttt{ffprobe}  & 384.5 & 14012.1 & 11728.4 & 19.5\% \\
\texttt{git}      & 67.5  & 31.5    & 23.1    & 36.4\% \\
\texttt{pdftops}  & 83.3  & 353.3   & 277.3   & 27.4\% \\
\texttt{cppcheck} & 71.1  & 676.3   & 612.6   & 10.4\% \\
\texttt{python}   & 553.1 & 205.2   & 178.3   & 15.1\% \\
\texttt{upx}      & 54.9  & 54.3    & 46.8    & 16.0\% \\
\texttt{h5dump}   & 877.3 & 668.2   & 513.5   & 30.1\% \\
\bottomrule
\end{tabular}
\end{table}

Table~\ref{tab:mfuzz-overhead} shows that {\mfuzz} consistently increases
peak memory usage relative to honggfuzz, but the additional cost remains
manageable for our empirical study.
The peak memory cost is dominated by the retained corpus, particularly the
number of preserved seeds and their sizes, whereas the in-memory profiling
and graph-related state of {\mfuzz} contributes a smaller fixed overhead for
driver attribution and coverage projection.
Accordingly, targets with larger or more rapidly growing corpora can exhibit
a more noticeable memory gap.
For example, on \texttt{ffprobe}, we observe a substantially larger peak
memory difference, which is consistent with a larger retained corpus:
{\mfuzz} preserves 193 more seeds, each averaging about 10\,MB, in addition
to the structural profiling overhead.
For smaller benchmarks, the relative increase can appear high while the
absolute cost remains low; for instance, \texttt{git} shows a 36.4\%
increase, but the absolute difference is only 8.4\,MB.
The switch-cost results show a similar pattern.
Larger targets generally incur higher switching overhead, which is expected
because profile snapshotting and coverage projection operate over larger
structural states.
Even the highest observed switch cost remains below 1\,s, and switching occurs
only at driver-slice boundaries rather than per test case, so this cost is
amortized over the entire execution slice.
Overall, these results indicate that {\mfuzz} preserves coverage behavior
under driver switching while introducing manageable runtime and memory
overhead on large representative targets from different benchmark domains.
This supports the use of {\mfuzz} as lightweight execution support for the
empirical study.

\section{Empirical Results} \label{sec:eval}

This section presents the empirical results for the study goals and research
questions in Section~\ref{sec:goal} using the methodology described in
Section~\ref{sec:method}.
All graph-based results are derived from driver-induced subgraphs projected onto
the whole-program call-graph backbone and are computed using the variables and
metrics defined in Section~\ref{ssec:metrics}.

\vspace{6pt}
\noindent
\textbf{Experimental Environment.}
All experiments were conducted on a 64-bit Ubuntu~18.04 workstation equipped
with a 24-core AMD Ryzen Threadripper 7960X CPU and 128~GB of RAM.
Each fuzzing instance was pinned to a dedicated CPU core and executed for
24~hours.
To reduce the effect of fuzzing nondeterminism, each experiment was repeated
five times.

\vspace{4pt}
\noindent
\textbf{Experimental Protocol.}
For each benchmark executable and its associated drivers, we run fuzzing under
the controlled driver-switching setup described in Section~\ref{sec:tool}, 
ensuring consistent instrumentation,
mutation policies, and runtime conditions across drivers.
The total fuzzing budget is divided evenly across drivers, and dynamic
profiles are synchronized only at driver-switch boundaries to preserve correct
per-driver attribution of function coverage and crash artifacts.
All structural measurements are computed offline from the collected profiles,
without introducing any feedback into the fuzzing process.
Since fuzzing is nondeterministic, repeated runs of the same executable may
produce slightly different coverage outcomes and induced structural subgraphs.
For scalar fuzzing outcomes, function coverage and crashes are attributed to
individual drivers and reported as averages across the five runs.
In contrast, CFG edge coverage is reported only at the executable level and is
not distinguished by driver.
For graph-based structural analysis, we aggregate repeated runs of the same
driver by taking the union of their covered functions and call edges on the
shared whole-program call graph.
All structural metrics are then computed on these aggregated driver-induced
subgraphs, yielding a more stable representation of the structure explored by
each driver.

\subsection{RQ1: Effectiveness of Fuzzing with Multi-driver}\label{sec:rq1}

RQ1 examines overall fuzzing effectiveness 
by comparing {\mfuzz} with {\mfuzz}-single (the best individual-driver configuration) under the same total
fuzzing budget. 
We evaluate coverage effectiveness using function-level coverage on the shared call-graph backbone 
and final CFG-edge coverage, 
and we additionally report bug-finding outcomes exposed by {\mfuzz}. 
This allows us to
assess whether combining drivers 
in {\mfuzz} improves structural exploration over the strongest single-driver configuration 
while also exposing practically meaningful failures.

\subsubsection{Coverage Effectiveness}

\begin{table*}[htp]
\centering
\small
\caption{Coverage results comparing {\mfuzz} and {\mfuzz}-single under the same
total fuzzing budget, in terms of covered call-graph nodes (\#CGNode) and
control-flow graph edges (\#CFGEdge). Delta shows the relative improvement of
{\mfuzz} over {\mfuzz}-single.}
\label{tab:rq1-cgnode-cfgedge-delta}
\resizebox{0.98\textwidth}{!}{
\begin{tabular}{llrrrrrrr}
\toprule
& & & \multicolumn{2}{c}{{\mfuzz}} & \multicolumn{2}{c}{{\mfuzz}-single} & \multicolumn{2}{c}{Delta} \\
\cmidrule(lr){4-5} \cmidrule(lr){6-7} \cmidrule(lr){8-9}
Benchmark & Executable & $\#Driver$ & $\#CGNode$ & $\#CFGEdge$ & $\#CGNode$ & $\#CFGEdge$ & $\#CGNode$ & $\#CFGEdge$ \\
\midrule

\multirow{2}{*}{snort3}
 & snort & 9 & 8394 & 9509 & 6359 & 7464 & 32.0\% & 27.4\% \\
 & snort2lua & 10 & 1924 & 4121 & 1884 & 3764 & 2.1\% & 9.5\% \\

\multirow{1}{*}{unbound}
 & unbound-checkconf & 7 & 62 & 340 & 61 & 271 & 1.6\% & 25.5\% \\

\multirow{2}{*}{http-parser}
 & parsertrace & 3 & 14 & 369 & 12 & 213 & 16.7\% & 73.2\% \\
 & url\_parser & 2 & 5 & 23 & 4 & 14 & 25.0\% & 64.3\% \\

\multirow{2}{*}{ffmpeg}
 & ffmpeg & 19 & 1423 & 5990 & 1178 & 3816 & 20.8\% & 57.0\% \\
 & ffprobe & 28 & 2415 & 13361 & 2273 & 12915 & 6.2\% & 3.5\% \\

\multirow{3}{*}{libtiff}
 & tiff2bw & 7 & 198 & 1275 & 163 & 576 & 21.5\% & 121.4\% \\
 & tiffinfo & 7 & 274 & 1804 & 271 & 1788 & 1.1\% & 0.9\% \\
 & tiff2pdf & 21 & 315 & 2233 & 295 & 1985 & 6.8\% & 12.5\% \\

\multirow{3}{*}{wavpack}
 & wavpack & 32 & 16 & 128 & 15 & 68 & 6.7\% & 88.2\% \\
 & wvunpack & 30 & 14 & 106 & 11 & 49 & 27.3\% & 116.3\% \\
 & wvgain & 12 & 15 & 71 & 12 & 41 & 25.0\% & 73.2\% \\

\multirow{1}{*}{git}
 & git & 6 & 222 & 549 & 216 & 536 & 2.8\% & 2.4\% \\

\multirow{3}{*}{sleuthkit}
 & istat & 4 & 74 & 89 & 66 & 77 & 12.1\% & 15.6\% \\
 & img\_stat & 7 & 174 & 129 & 168 & 111 & 3.6\% & 16.2\% \\
 & tsk\_recover & 4 & 74 & 89 & 66 & 77 & 12.1\% & 15.6\% \\

\multirow{1}{*}{file}
 & file & 12 & 176 & 1468 & 134 & 483 & 31.3\% & 203.9\% \\

\multirow{3}{*}{xpdf}
 & pdfdetach & 12 & 317 & 1310 & 322 & 1356 & -1.6\% & -3.4\% \\
 & pdfinfo & 25 & 368 & 1572 & 354 & 1318 & 4.0\% & 19.3\% \\
 & pdftops & 66 & 744 & 2947 & 670 & 2228 & 11.0\% & 32.3\% \\

\multirow{1}{*}{libxml2}
 & xmllint & 40 & 657 & 5586 & 499 & 3570 & 31.7\% & 56.5\% \\

\multirow{1}{*}{jq}
 & jq & 18 & 390 & 1876 & 354 & 1554 & 10.2\% & 20.7\% \\

\multirow{3}{*}{binutils}
 & objdump & 22 & 919 & 7885 & 627 & 3906 & 46.6\% & 101.9\% \\
 & readelf & 29 & 301 & 4595 & 184 & 2366 & 63.6\% & 94.2\% \\
 & addr2line & 8 & 392 & 2572 & 368 & 2265 & 6.5\% & 13.6\% \\

\multirow{1}{*}{cppcheck}
 & cppcheck & 48 & 1763 & 4526 & 1745 & 4098 & 1.0\% & 10.4\% \\

\multirow{1}{*}{libdwarf}
 & dwarfdump & 29 & 376 & 1294 & 324 & 1232 & 16.0\% & 5.0\% \\

\multirow{1}{*}{cpython3}
 & python & 23 & 3948 & 21062 & 3652 & 18604 & 8.1\% & 13.2\% \\

\multirow{2}{*}{quickjs}
 & qjs & 6 & 944 & 8398 & 689 & 1784 & 37.0\% & 370.7\% \\
 & qjsc & 27 & 639 & 6966 & 637 & 6943 & 0.3\% & 0.3\% \\

\multirow{1}{*}{lua}
 & lua & 36 & 985 & 4957 & 250 & 563 & 294.0\% & 780.5\% \\

\multirow{2}{*}{libarchive}
 & bsdtar & 15 & 493 & 2129 & 502 & 3160 & -1.8\% & -32.6\% \\
 & bsdunzip & 13 & 240 & 1394 & 176 & 888 & 36.4\% & 57.0\% \\

\multirow{1}{*}{upx}
 & upx & 22 & 1914 & 5019 & 1914 & 4973 & 0.0\% & 0.9\% \\

\multirow{1}{*}{xz}
 & xz & 17 & 232 & 789 & 216 & 246 & 7.4\% & 220.7\% \\

\multirow{3}{*}{hdf5}
 & h5dump & 12 & 1860 & 8472 & 1257 & 5407 & 48.0\% & 56.7\% \\
 & h5ls & 12 & 1765 & 7998 & 1676 & 7656 & 5.3\% & 4.5\% \\
 & h5repack & 37 & 2298 & 10076 & 2183 & 9608 & 5.3\% & 4.9\% \\

\multirow{3}{*}{netcdf}
 & ncdump & 17 & 443 & 1922 & 369 & 1388 & 20.1\% & 38.5\% \\
 & ncgen & 22 & 248 & 348 & 66 & 120 & 275.8\% & 190.0\% \\
 & nccopy & 30 & 17 & 67 & 14 & 24 & 21.4\% & 179.2\% \\

\multirow{1}{*}{sqlite3}
 & sqlite3 & 48 & 392 & 2536 & 392 & 2536 & 0.0\% & 0.0\% \\

\midrule
\multicolumn{7}{r}{Average} & 27.9\% & 73.5\% \\
\bottomrule
\end{tabular}}
\end{table*}

Table~\ref{tab:rq1-cgnode-cfgedge-delta} compares {\mfuzz} with {\mfuzz}-single 
under the same total fuzzing budget. 
Effectiveness is evaluated
using two structural coverage metrics: covered call-graph nodes (\#CGNode) and
covered control-flow graph edges (\#CFGEdge). 
Overall, 
{\mfuzz} improves
structural exploration for most executables. 
Across the benchmark suite, 
it increases coverage by an average of \textbf{27.9\%} in \#CGNode and \textbf{73.5\%} in \#CFGEdge.

For most executables, improvements are observed in both \#CGNode and \#CFGEdge,
indicating that multi-driver fuzzing not only reaches more functions
but also explores more intra-procedural control flow within the reached regions.
Representative examples include \texttt{ffmpeg} (\(+20.8\%\) in \#CGNode, \(+57.0\%\) in \#CFGEdge),
\texttt{libxml2/xmllint} (\(+31.7\%, +56.5\%\)),
\texttt{cpython3/python} (\(+8.1\%, +13.2\%\)),
and \texttt{netcdf/ncdump} (\(+20.1\%, +38.5\%\)).
These consistent improvements suggest that different drivers expose complementary execution regions
rather than repeatedly exercising the same structural areas.
The gains are often substantially larger for \#CFGEdge than for \#CGNode.
This pattern can be seen in \texttt{file} (\(+31.3\%\) in \#CGNode, \(+203.9\%\) in \#CFGEdge),
\texttt{quickjs/qjs} (\(+37.0\%, +370.7\%\)),
\texttt{xz} (\(+7.4\%, +220.7\%\)),
and \texttt{binutils/objdump} (\(+46.6\%, +101.9\%\)).
Such cases indicate that multi-driver fuzzing not only expands the set of reached functions,
but also deepens exploration within already reached regions by activating additional control-flow alternatives.
This asymmetry is important because it shows that driver diversity can improve both
\emph{breadth} and \emph{depth} of structural exploration.
Several executables exhibit especially large gains, indicating strong complementarity among drivers.
For example, \texttt{lua} improves by \textbf{\(+294.0\%\)} in \#CGNode
and \textbf{\(+780.5\%\)} in \#CFGEdge,
while \texttt{netcdf/ncgen} improves by \textbf{\(+275.8\%\)} and \textbf{\(+190.0\%\)}.
Similarly, \texttt{hdf5/h5dump} shows gains of \((+48.0\%, +56.7\%)\),
and \texttt{binutils/readelf} improves by \((+63.6\%, +94.2\%)\).
These cases suggest that single-driver fuzzing leaves substantial executable-specific regions unexplored,
whereas combining multiple drivers covers substantially larger portions of the software structure.

The magnitude of improvement is nevertheless uneven across executables.
Some cases show only marginal gains,
including \texttt{cppcheck/cppcheck} (\(+1.0\%, +10.4\%\)),
\texttt{quickjs/qjsc} (\(+0.3\%, +0.3\%\)),
\texttt{git} (\(+2.8\%, +2.4\%\)),
and \texttt{libtiff/tiffinfo} (\(+1.1\%, +0.9\%\)).
These small deltas suggest that the corresponding drivers are likely to be highly overlapping
or otherwise structurally redundant,
so additional drivers contribute little new structure beyond the strongest individual one.
This interpretation is consistent with the low-dispersion cases reported in RQ2 (Section~\ref{sec:rq2})
and the higher-overlap patterns analyzed in RQ4 (Section~\ref{sec:rq4}).
A small number of executables show slight regressions under multi-driver fuzzing,
notably \texttt{xpdf/pdfdetach} (\(-1.6\%, -3.4\%\))
and \texttt{libarchive/bsdtar} (\(-1.8\%, -32.6\%\)).
These are exceptions rather than the dominant trend.
Under a fixed total budget, distributing effort across multiple drivers can reduce
the time allocated to a particularly effective single driver,
especially when drivers overlap heavily or when one driver dominates exploration.
This behavior is consistent with the driver-imbalance results in RQ2.
Near-parity appears only in a few cases,
such as \texttt{sqlite3} (\(0.0\%, 0.0\%\)) and \texttt{upx} (\(0.0\%, +0.9\%\)).
These results suggest that some programs have highly shared execution backbones across drivers,
so multi-driver fuzzing offers little additional structural benefit.
As shown later in RQ2 and RQ4,
such cases are associated with low driver heterogeneity and high structural redundancy.

\subsubsection{Bug Discovery}

Table~\ref{tab:bug-findings} shows that 
the structural gains of {\mfuzz} also lead to practically meaningful defect-finding results. 
In total, 
{\mfuzz} exposed \textbf{13} unique bugs and abnormal behaviors across four benchmarks.
These findings are concentrated 
in a small set of executables and driver combinations, 
suggesting that bug exposure is strongly dependent on execution mode. 
Among the 13 findings, 
only \textbf{3} are also marked as reachable by {\mfuzz}-single, 
while the remaining majority are exposed only when multiple drivers are combined in {\mfuzz}. 
The findings also span different severity and maturity levels, 
including newly submitted bugs in \texttt{ncdump}, confirmed known abnormal behaviors in \texttt{h5repack} and \texttt{readelf}, 
and a pending hang in \texttt{lua}. 
Overall, 
this shows that the benefit of {\mfuzz} is not limited to improving aggregate structural coverage, 
but also includes exposing failure behaviors that remain hidden under the strongest
single-driver configuration.

The findings are also closely tied to 
semantically specialized execution modes rather than generic default behavior. 
In \texttt{ncdump}, 
the exposed bugs are triggered by formatting and extraction-related options. 
In \texttt{h5repack},
they cluster around ordering and transformation modes. 
In \texttt{readelf},
they arise under debug-display and whole-file reporting modes, 
while in \texttt{lua} the hang appears under the \texttt{-E} execution mode. 
These are not merely arbitrary command-line variations. 
Instead, they activate different semantic slices of the same executable and 
therefore exercise different parsing, output, transformation, cleanup, and error-handling paths. 
This helps explain why many findings are not recovered by {\mfuzz}-single 
even when that configuration is the strongest one overall: 
higher aggregate coverage does not necessarily provide access to the specific semantic modes 
in which these failures occur.

Taken together, 
these results indicate that bug discovery in configuration-rich executables 
is driver-sensitive in much the same way as structural exploration. 
They therefore reinforce the main conclusion of RQ1:
combining drivers in {\mfuzz} is beneficial not only because it broadens structural reach, 
but also because it exposes semantically distinct execution behaviors that can uncover bugs and 
abnormal states missed by the best single-driver configuration. 
More broadly, 
the results suggest that driver-aware fuzzing 
should preserve execution-mode diversity rather than collapsing effort 
onto a single dominant driver, since rare or specialized modes may contribute disproportionately 
to bug exposure despite accounting for only a limited share of overall coverage.

\begin{table*}[t]
\centering
\small
\caption{Unique bugs and abnormal behaviors exposed by {\mfuzz}, totaling 11 findings. 
Findings marked with $\dagger$ were also exposed by {\mfuzz}-single, totaling 3. Detailed per-bug reports are provided in the artifact.}
\label{tab:bug-findings}
\begin{tabular}{llp{2.8cm}p{4.2cm}p{1.1cm}p{1.8cm}}
\toprule
Benchmark & Executable & Triggering driver(s) & Exposed bug(s) & Known$?$ & Status \\
\midrule
Netcdf & \texttt{ncdump} &
\makecell[l]{\texttt{-i}\\ \texttt{-b f}} &
1$\times$ Assertion; 2 $\times$ Heap corruptions &
Unknown & Issue submitted \\

Netcdf & \texttt{ncdump} &
\makecell[l]{\texttt{-c}\\ \texttt{-x}} &
1 $\times$ Double free$^{\dagger}$ &
Unknown & Issue submitted \\

HDF5 & \texttt{h5repack} &
\makecell[l]{\texttt{-n}\\ \texttt{-q name}\\ \texttt{-q creation\_order}\\ \texttt{-z ascending}} &
2 $\times$ Double Free; 2 $\times$ Heap Corruption &
Known & Confirmed \\

Binutils & \texttt{readelf} &
\texttt{-w all} &
1 $\times$ SIGABRT &
Known & Confirmed \\

Binutils & \texttt{readelf} &
\texttt{-a} &
1 $\times$ SIGABRT$^{\dagger}$ &
Known & Confirmed \\

Lua & \texttt{lua} &
\texttt{-E} &
1 $\times$ Hang$^{\dagger}$ &
Unknown & Pending \\
\bottomrule
\end{tabular}
\end{table*}

\find{
{\mfuzz} is generally more effective than {\mfuzz}-single under the same total
fuzzing budget. Across the benchmark suite, it improves coverage by an average
of \textbf{27.9\%} in call-graph nodes and \textbf{73.5\%} in control-flow
graph edges, showing gains in both the breadth and depth of structural
exploration. Beyond coverage, {\mfuzz} also exposed \textbf{13} unique bugs and
abnormal behaviors, only \textbf{3} of which were also found by
{\mfuzz}-single. These results show that combining drivers in {\mfuzz}
improves not only structural reachability, but also practical bug discovery by
exercising semantic behaviors that the strongest single-driver configuration
does not reliably reach.
}

\subsection{RQ2: Driver Effectiveness Under Equal Time Budgets}\label{sec:rq2}

\begin{table*}[htp]
\centering
\small
\caption{Dispersion of normalized per-driver call-graph node coverage under equal time budgets.
$\#Driver$ denotes the number of drivers, and 
$\#CGNode$ denotes the overlap-aware union of covered call-graph nodes across all drivers.
Std, CV, and Min/Median/Max summarize the distribution of normalized per-driver coverage shares.}
\label{tab:driver-dispersion-cg-node}
\resizebox{0.86\textwidth}{!}{
\begin{tabular}{llrrrrrrr}
\toprule
Benchmark & Executable & $\#Driver$ & $\#CGNode$ & Std & CV & Min & Median & Max \\
\midrule

\multirow{2}{*}{snort3}
 & snort & 9 & 8394 & 0.0786 & 0.1133 & 0.5825 & 0.6748 & 0.8764 \\
 & snort2lua & 10 & 1924 & 0.0433 & 0.0446 & 0.8540 & 0.9923 & 1.0000 \\

\multirow{1}{*}{unbound}
 & checkconf & 7 & 62 & 0.3386 & 0.3930 & 0.0323 & 1.0000 & 1.0000 \\

\multirow{2}{*}{http-parser}
 & parsertrace & 3 & 14 & 0.0673 & 0.0707 & 0.8571 & 1.0000 & 1.0000 \\
 & url\_parser & 2 & 5 & 0.1000 & 0.1111 & 0.8000 & 0.9000 & 1.0000 \\

\multirow{2}{*}{ffmpeg}
 & ffmpeg & 19 & 1423 & 0.3388 & 1.6137 & 0.0632 & 0.0632 & 0.9958 \\
 & ffprobe & 28 & 2415 & 0.3812 & 1.2437 & 0.0207 & 0.0253 & 0.9213 \\

\multirow{3}{*}{libtiff}
 & tiff2bw & 7 & 198 & 0.4774 & 0.8354 & 0.0152 & 0.9848 & 0.9848 \\
 & tiffinfo & 7 & 274 & 0.0866 & 0.1051 & 0.7664 & 0.7701 & 0.9818 \\
 & tiff2pdf & 21 & 315 & 0.0112 & 0.0119 & 0.9206 & 0.9397 & 0.9778 \\

\multirow{3}{*}{wavpack}
 & wavpack & 32 & 16 & 0.1954 & 0.2387 & 0.5000 & 0.9375 & 1.0000 \\
 & wvunpack & 30 & 14 & 0.0878 & 0.1142 & 0.4286 & 0.7857 & 0.8571 \\
 & wvgain & 12 & 15 & 0.1499 & 0.1984 & 0.2667 & 0.8000 & 0.8667 \\

\multirow{1}{*}{git}
 & git & 6 & 222 & 0.1528 & 0.1689 & 0.5631 & 0.9730 & 0.9730 \\

\multirow{3}{*}{sleuthkit}
 & istat & 4 & 74 & 0.0147 & 0.0161 & 0.8919 & 0.9189 & 0.9324 \\
 & img\_stat & 7 & 174 & 0.4634 & 0.8275 & 0.0172 & 0.9598 & 0.9655 \\
 & tsk\_recover & 4 & 74 & 0.0147 & 0.0161 & 0.8919 & 0.9189 & 0.9324 \\

\multirow{1}{*}{file}
 & file & 12 & 176 & 0.2228 & 0.7295 & 0.1591 & 0.2216 & 0.9489 \\

\multirow{3}{*}{xpdf}
 & pdfdetach & 12 & 317 & 0.3639 & 2.0031 & 0.0189 & 0.0189 & 0.9968 \\
 & pdfinfo & 25 & 368 & 0.3480 & 2.0364 & 0.0190 & 0.0190 & 0.9864 \\
 & pdftops & 66 & 744 & 0.3418 & 2.2620 & 0.0067 & 0.0067 & 0.9772 \\

\multirow{1}{*}{libxml2}
 & xmllint & 40 & 657 & 0.2327 & 0.5020 & 0.0091 & 0.5495 & 0.7458 \\

\multirow{1}{*}{jq}
 & jq & 18 & 390 & 0.3193 & 0.8563 & 0.0462 & 0.3051 & 0.9564 \\

\multirow{3}{*}{binutils}
 & objdump & 22 & 919 & 0.1527 & 0.3535 & 0.1055 & 0.3645 & 0.8020 \\
 & readelf & 29 & 301 & 0.1381 & 0.4129 & 0.2492 & 0.2757 & 0.8339 \\
 & addr2line & 8 & 392 & 0.4591 & 0.7310 & 0.0353 & 0.9810 & 0.9891 \\

\multirow{1}{*}{cppcheck}
 & cppcheck & 48 & 1763 & 0.0657 & 0.0748 & 0.8395 & 0.8395 & 0.9977 \\

\multirow{1}{*}{libdwarf}
 & dwarfdump & 29 & 376 & 0.0456 & 0.0638 & 0.6782 & 0.7021 & 0.8723 \\

\multirow{1}{*}{cpython3}
 & python & 23 & 3948 & 0.2889 & 0.3792 & 0.0251 & 0.8528 & 0.9620 \\

\multirow{2}{*}{quickjs}
 & qjs & 6 & 944 & 0.2281 & 0.2686 & 0.3485 & 0.9550 & 0.9894 \\
 & qjsc & 27 & 639 & 0.3087 & 2.7118 & 0.0047 & 0.0047 & 0.9937 \\

\multirow{1}{*}{lua}
 & lua & 36 & 985 & 0.2479 & 0.5783 & 0.2538 & 0.3066 & 0.9939 \\

\multirow{2}{*}{libarchive}
 & bsdtar & 15 & 493 & 0.2867 & 2.1396 & 0.0203 & 0.0203 & 0.9026 \\
 & bsdunzip & 13 & 240 & 0.0851 & 0.0951 & 0.7417 & 0.9375 & 0.9583 \\

\multirow{1}{*}{upx}
 & upx & 22 & 1914 & 0.0006 & 0.0006 & 0.9974 & 0.9990 & 1.0000 \\

\multirow{1}{*}{xz}
 & xz & 17 & 232 & 0.0901 & 0.2159 & 0.2269 & 0.3935 & 0.6204 \\

\multirow{3}{*}{hdf5}
 & h5dump & 12 & 1860 & 0.0981 & 0.1065 & 0.6742 & 0.9710 & 0.9871 \\
 & h5ls & 12 & 1765 & 0.0828 & 0.1108 & 0.5711 & 0.7331 & 0.9416 \\
 & h5repack & 37 & 2298 & 0.3356 & 0.6464 & 0.2215 & 0.2293 & 0.9386 \\

\multirow{3}{*}{netcdf}
 & ncdump & 17 & 443 & 0.3396 & 0.6099 & 0.0045 & 0.6456 & 0.9165 \\
 & ncgen & 22 & 248 & 0.4149 & 0.6731 & 0.1212 & 0.9697 & 1.0000 \\
 & nccopy & 30 & 17 & 0.0957 & 0.3938 & 0.1765 & 0.2353 & 0.6471 \\

\multirow{1}{*}{sqlite3}
 & sqlite3 & 48 & 392 & 0.0000 & 0.0000 & 1.0000 & 1.0000 & 1.0000 \\

\bottomrule
\end{tabular}}
\end{table*}

To evaluate driver effectiveness under equal time budgets,
we use call-graph node coverage (\#CGNode) as a measure of function-level structural activation.
For each executable, a driver’s effectiveness is represented by its normalized coverage share
with respect to the overlap-aware union of covered call-graph nodes across all drivers.
Under identical fuzzing time for each driver,
Table~\ref{tab:driver-dispersion-cg-node} summarizes the distribution of these
\emph{normalized per-driver} coverage shares using standard deviation (Std),
coefficient of variation (CV), and Min/Median/Max values.

Across the benchmark suite,
driver effectiveness under equal time budgets exhibits substantial dispersion and wide variability.
The coefficient of variation (CV) ranges from $0$ to above $2.7$.
For example, \texttt{sqlite3} shows CV $=0$, indicating perfectly uniform contribution across its 48 drivers,
whereas \texttt{qjsc} exhibits CV $=2.71$, reflecting extreme imbalance.
This wide range shows that equal scheduling does not systematically produce equal structural contribution.
A recurring pattern is pronounced right-skewness in the distribution of effectiveness shares.
In many executables, minimum shares approach zero while maximum shares approach one.
For example, \texttt{pdftops} (66 drivers) has Min $=0.0067$ and Max $=0.9772$,
while \texttt{ffmpeg} has Min $=0.0632$ and Max $=0.9958$.
Median values are often substantially lower than the maximum,
indicating that structural activation is concentrated in a small subset of drivers.
In extreme cases, one driver dominates exploration while several others contribute negligibly.
This dominant-driver pattern is especially visible in \texttt{qjsc} and \texttt{bsdtar},
where the maximum share is close to $1$ but the median remains close to zero.

In contrast,
near-uniform contribution is rare and appears only in a small subset of executables,
such as \texttt{sqlite3} and \texttt{upx},
where Min/Median/Max values are tightly clustered around $1$ and CV is nearly zero.
In these cases, drivers activate almost identical structural regions,
suggesting that configuration options largely share a common execution backbone.
This observation is consistent with the near-parity cases in RQ1 (Section~\ref{sec:rq1}),
where multi-driver fuzzing provides little additional benefit over the best single driver.
Most executables lie between these extremes and exhibit moderate dispersion.
For example, \texttt{wavpack} (CV $=0.24$) and \texttt{h5ls} (CV $=0.11$)
show noticeable but not dominant imbalance.
In such cases, drivers differ meaningfully in effectiveness,
but no single driver fully dominates exploration.
Even in these intermediate cases,
structural contribution remains uneven,
indicating that heterogeneity is not confined to a few extreme outliers
but is a systematic property of driver-based execution.
This dispersion pattern appears across multiple software domains,
including media tools (\texttt{ffmpeg}), document processors (\texttt{xpdf}),
language runtimes (\texttt{quickjs}), and storage systems (\texttt{sqlite3}).
The consistency of this pattern across domains suggests that effectiveness heterogeneity
is not specific to a particular workload class,
but reflects a broader structural property of configuration-driven software.

These results also help explain the effectiveness differences observed in RQ1.
Executables with highly imbalanced driver contributions are more likely to benefit from
multi-driver fuzzing when distinct drivers expose complementary structural regions,
but they may also suffer from redundant budget allocation when many weak drivers overlap
with a dominant one.
Conversely, near-uniform cases such as \texttt{sqlite3} and \texttt{upx}
indicate strong redundancy across drivers, which explains why multi-driver fuzzing
offers little additional structural gain in those executables.

\find{
Driver effectiveness under equal time budgets is systematically heterogeneous.
While a small subset of programs exhibits near-uniform contribution,
most executables demonstrate moderate to strong imbalance,
and several show dominant-driver behavior in which a few drivers account for most structural activation.
Equal resource allocation therefore does not guarantee balanced exploration across configuration space,
which helps explain why round-robin multi-driver fuzzing can be effective in some executables
but inefficient in others.
}

\subsection{RQ3: Structural Organization and Modularity} \label{sec:rq3}

\begin{figure}[htp]
  \centering
  \includegraphics[width=1\textwidth]{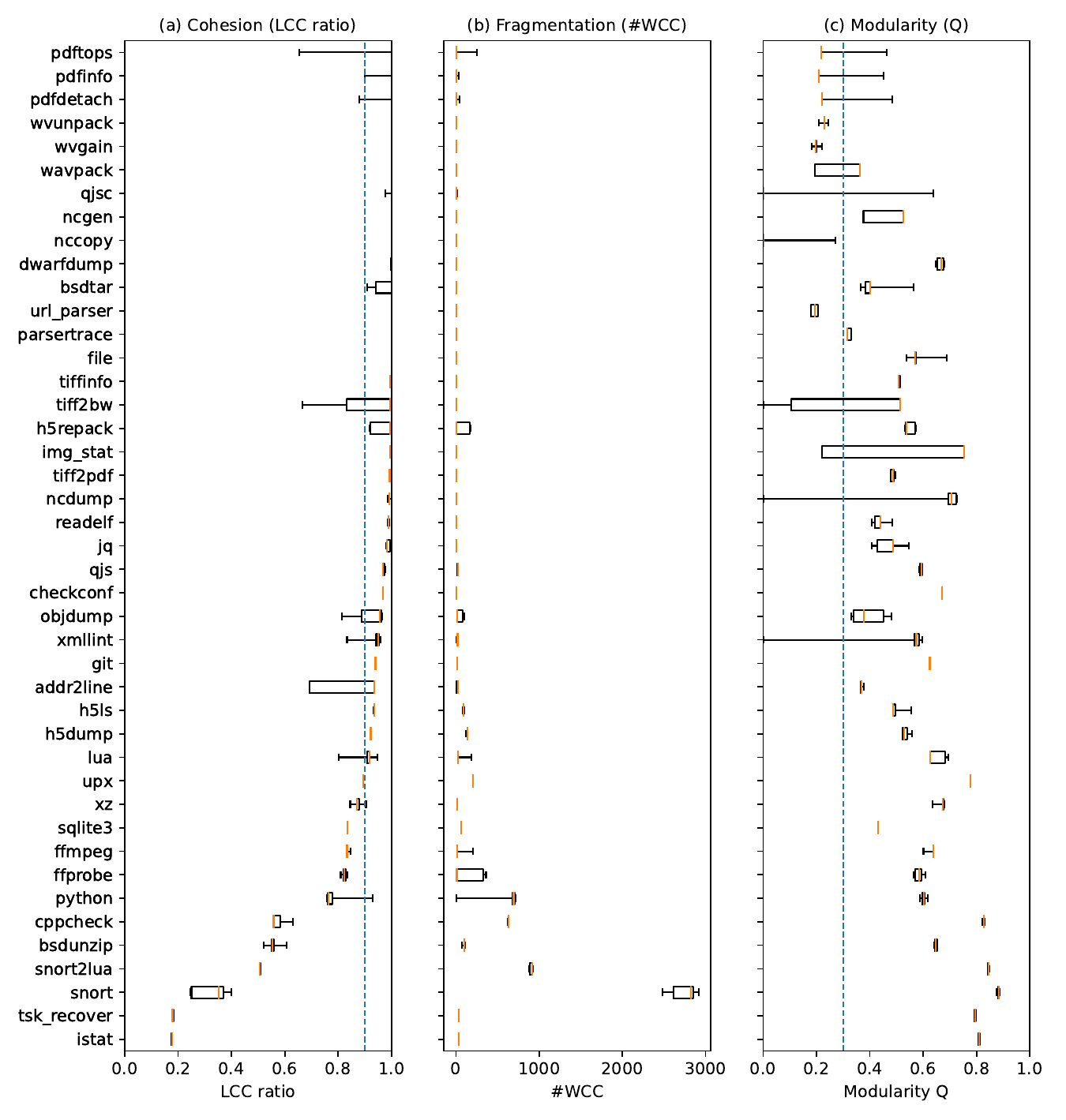}
  \caption{Structural organization of driver-induced subgraphs across executables.
            Boxplots summarize per-driver distributions for each executable.
            (a) Largest Connected Component (LCC) ratio measures cohesion.
            (b) Number of Weakly Connected Components (\#WCC) measures fragmentation.
            (c) Modularity ($Q$) measures community structure.
            Executables are ordered by median LCC ratio.
            The results show substantial structural variation across drivers for many programs
            (e.g., \texttt{ffmpeg}, \texttt{quickjs}, \texttt{libxml2}),
            while others (e.g., \texttt{sqlite3}, \texttt{upx}) remain consistently cohesive,
            indicating that configuration modes can either concentrate execution within unified regions
            or activate structurally distinct subsystems.}
  \label{fig:driver-sharing}
\end{figure}

We answer RQ3 by analyzing the structural organization of driver-induced subgraphs
within the whole-program call graph.
All measurements are performed on driver-induced subgraphs using the shared whole-program
call graph as a common structural reference, enabling direct comparison across drivers
and executables.
For each driver, we evaluate three graph-level properties:
Largest Connected Component (LCC) ratio for cohesion,
Number of Weakly Connected Components (\#WCC) for fragmentation,
and modularity ($Q$) for community structure.
Figure~\ref{fig:driver-sharing} summarizes the per-driver distributions of these metrics across executables.

\subsubsection{Cohesion (LCC Ratio)}
Figure~\ref{fig:driver-sharing}(a) reports the per-driver distribution of LCC ratios across executables.
Across the benchmark suite, LCC ratios span a wide range, from approximately $0.15$ to $1.0$.
Many executables exhibit substantial within-executable variation,
with some drivers concentrated near a single cohesive region and others spread across multiple structural regions.
For example, \texttt{snort} includes drivers with LCC ratios below $0.3$ and others above $0.9$,
while \texttt{ffmpeg} spans roughly $0.3$ to near $1.0$.
In contrast, 
\texttt{sqlite3} and \texttt{upx} remain tightly concentrated near $1.0$ across drivers.
These results indicate that configuration modes often activate structurally distinct regions within the same executable:
high-LCC drivers remain concentrated in a dominant connected region,
whereas low-LCC drivers are distributed across multiple weaker components.

This variation is consistent with the effectiveness differences observed in RQ1 (Section~\ref{sec:rq1}) and RQ2 (Section~\ref{sec:rq2}).
Executables with consistently high LCC ratios across drivers, 
such as \texttt{sqlite3} and \texttt{upx}, also showed near-parity in RQ1 and near-uniform contribution in RQ2,
suggesting that their drivers largely traverse a shared structural core.
By contrast, executables with wide LCC dispersion, 
such as \texttt{ffmpeg} and \texttt{snort},
are more consistent with heterogeneous driver behavior.

\find{
Driver-induced subgraphs exhibit substantial variation in structural cohesion.
While a small subset of executables remains consistently cohesive across drivers,
most programs show wide dispersion in LCC ratios,
indicating that configuration modes often activate structurally distinct execution regions within the same executable.
}

\subsubsection{Fragmentation (\#WCC)}
Figure~\ref{fig:driver-sharing}(b) shows the per-driver distribution of weakly connected component counts (\#WCC).
Across the benchmark suite, \#WCC ranges from $1$ to above $20$.
Many executables show strong within-executable variation:
some drivers induce a single connected component,
whereas others activate many disconnected structural regions.
For example, \texttt{snort} includes drivers whose \#WCC exceeds $15$,
while others remain near $1$.
Similarly, \texttt{ffmpeg} and \texttt{python} show drivers with markedly larger component counts than their medians.
In contrast, \texttt{sqlite3} and \texttt{upx} remain tightly concentrated at \#WCC $=1$ across drivers.

These results complement the LCC analysis.
Drivers with higher fragmentation tend to activate multiple structurally disjoint subsystems,
whereas low-\#WCC drivers remain confined to a single cohesive region.
This supports the view that configuration options can selectively enable independent feature modules
or specialized processing pipelines.
It also helps explain why equal driver budgets in RQ2 (Section~\ref{sec:rq2}) do not imply equal structural contribution:
drivers that reach more fragmented regions may expose structural areas that other drivers do not cover.

\find{
Driver-induced subgraphs exhibit substantial variation in structural fragmentation.
While some executables maintain consistently connected subgraphs,
many show heavy-tailed dispersion in \#WCC,
indicating that certain drivers activate multiple disjoint structural regions.
Configuration modes therefore differ substantially in how they partition execution structure.
}

\subsubsection{Modularity ($Q$)}
Figure~\ref{fig:driver-sharing}(c) presents the per-driver distribution of modularity scores ($Q$).
Across the benchmark suite, modularity values range from near $0$ to above $0.6$.
Many executables show wide within-executable dispersion,
with some drivers exhibiting weak community structure and others showing much stronger separation into structural clusters.
For example, \texttt{ffmpeg} includes drivers with $Q$ near $0.1$ and others above $0.5$.
Similarly, \texttt{qjsc} and \texttt{pdftops} exhibit broad modularity ranges across drivers.
In contrast, \texttt{sqlite3} and \texttt{upx} remain tightly clustered at low $Q$ values.

Higher modularity is consistent with drivers activating more clearly separated functional communities,
whereas lower modularity indicates more homogeneous structural interaction.
Together with the LCC and \#WCC results, this shows that driver-induced subgraphs vary not only in size or overlap,
but also in how their internal structure is organized.
This structural variation provides a graph-level explanation for the heterogeneous effectiveness observed in RQ2 (Section~\ref{sec:rq2})
and anticipates the overlap patterns analyzed in RQ4 (Section~\ref{sec:rq4}).

\find{
Driver-induced subgraphs differ substantially in community organization.
While some executables exhibit consistently low modularity,
many show wide dispersion in $Q$ values,
indicating that configuration modes can selectively activate distinct functional communities.
Structural organization therefore varies systematically across drivers.
}

\subsection{RQ4: Driver Overlap and Redundancy} \label{sec:rq4}

\begin{figure}[htp]
  \centering
  \includegraphics[width=0.9\textwidth]{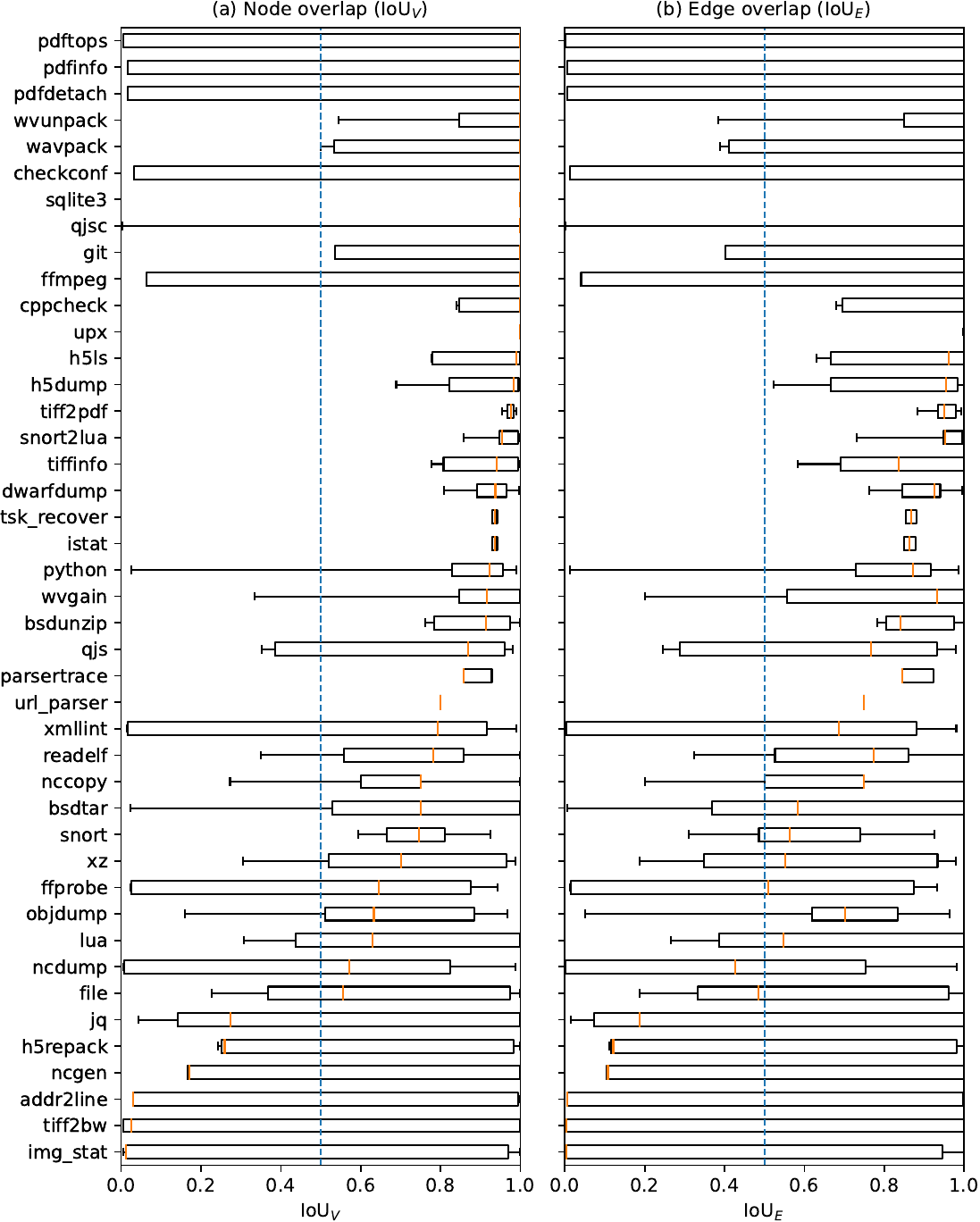}
  \caption{Distribution of pairwise structural overlap among driver-induced subgraphs.
  Each horizontal boxplot summarizes intersection-over-union (IoU) values computed
  between all driver pairs within the same executable.
  Panel (a) reports node-level overlap (IoU$_V$), and panel (b) reports edge-level overlap (IoU$_E$).
  Executables are ordered by median IoU$_V$ in ascending order.
  The dashed vertical line at 0.5 indicates moderate structural overlap.}
  \label{fig:rq4-iou}
\end{figure}

To evaluate structural redundancy among execution modes,
we measure pairwise structural overlap between driver-induced subgraphs within the same executable.
Overlap is quantified using intersection-over-union (IoU) over activated call-graph node sets (IoU$_V$)
and activated call-graph edge sets (IoU$_E$).
Each executable is therefore characterized by the distribution of IoU values
computed across all driver pairs.
Figure~\ref{fig:rq4-iou} summarizes these per-executable distributions,
ordered by median IoU$_V$ in ascending order.

Across the benchmark suite,
structural overlap spans nearly the full range from 0 to 1.
Some executables exhibit consistently low pairwise overlap,
indicating that different drivers activate largely distinct structural regions.
For example, \texttt{img\_stat}, \texttt{addr2line}, \texttt{tiff2bw}, and \texttt{ncdump}
show median IoU$_V$ values close to zero,
with most driver pairs sharing little of the call graph.
These cases suggest that the corresponding configuration modes map to relatively independent subsystems.
At the other extreme, executables such as \texttt{snort2lua}, \texttt{tiffinfo}, \texttt{h5dump}, and \texttt{upx}
show median IoU$_V$ values above $0.8$ with compact interquartile ranges near $1$.
In these programs, different drivers activate highly similar structural regions,
indicating substantial redundancy across execution modes.

Most executables lie between these extremes and exhibit moderate overlap with substantial dispersion.
Examples include \texttt{ffmpeg}, \texttt{readelf}, \texttt{objdump}, \texttt{jq}, and \texttt{python},
whose median IoU$_V$ values fall roughly between $0.4$ and $0.7$ and whose interquartile ranges remain wide.
These distributions indicate that redundant and orthogonal driver pairs often coexist within the same executable.
Some drivers traverse highly similar structural regions,
whereas others activate relatively distinct portions of the call graph.
This mixed-overlap pattern suggests that configuration spaces often contain multiple structural families
rather than forming either a purely redundant or a purely orthogonal set of execution modes.
Several executables show especially broad overlap ranges,
with IoU values extending from near zero to near one.
For example, \texttt{pdftops} and \texttt{xmllint} span almost the full overlap range across driver pairs.
Such cases suggest that configuration spaces can decompose into groups of closely related drivers
that coexist with more specialized drivers activating orthogonal subsystems.
This clustering behavior provides a structural explanation for the heterogeneous driver effectiveness observed in RQ2
and helps explain why multi-driver fuzzing in RQ1 can yield large gains even when some drivers are individually redundant.

Node-level and edge-level overlap follow broadly consistent trends,
although edge overlap (IoU$_E$) is often more concentrated near the upper bound.
In highly redundant executables such as \texttt{snort2lua} and \texttt{upx},
IoU$_E$ frequently approaches $1$,
indicating that drivers not only activate similar function sets
but also preserve nearly identical caller--callee relationships.
By contrast, executables such as \texttt{ffmpeg} and \texttt{python}
often show moderate IoU$_V$ together with wider dispersion in IoU$_E$.
This suggests that drivers may activate overlapping functional regions
while still inducing different interprocedural connectivity patterns within those regions.
Configuration therefore influences not only which functions are activated,
but also how those functions are structurally connected.

These overlap patterns are consistent with the earlier RQs.
Executables with high redundancy, such as \texttt{sqlite3} and \texttt{upx},
also showed near-uniform driver contribution in RQ2 and little improvement in RQ1,
indicating that multiple drivers largely traverse the same shared structural core.
Conversely, executables with broader overlap dispersion, such as \texttt{ffmpeg}, \texttt{objdump}, and \texttt{python},
are more consistent with the heterogeneous effectiveness patterns in RQ2
and the larger multi-driver gains observed in RQ1.
The consistency of these patterns across media tools, document processors,
toolchain utilities, language runtimes, and storage systems
suggests that structural redundancy is not workload-specific,
but a broader property of configuration-driven software.

\find{
Driver-induced subgraphs exhibit highly heterogeneous structural overlap.
Some executables show strong structural diversity, where most driver pairs activate largely disjoint call-graph regions,
while others show substantial redundancy, with most drivers traversing nearly identical structural cores.
Many programs also contain mixed overlap structure, where highly overlapping and weakly overlapping driver pairs coexist within the same executable.
Node-level and edge-level overlap follow consistent overall trends,
with edge overlap often more tightly concentrated near the upper bound.
Structural overlap therefore varies systematically across programs
and provides a graph-level explanation for both the effectiveness heterogeneity in RQ2
and the multi-driver gains observed in RQ1.
}
\subsection{RQ5: Structural Under-Exploration in Multi-Driver Fuzzing}\label{sec:rq5}

We next examine which structural regions remain under-explored after round-robin multi-driver fuzzing.
Our analysis proceeds in two stages.
First, we summarize residual region coverage over the shared structural backbone of each executable and group executables into recurring residual-gap categories based on the severity and shape of their remaining coverage gaps.
We then use representative case studies from each category to explain why these regions remain difficult to explore.
Table~\ref{tab:rq5-residual-region-coverage} reports the per-executable residual-coverage summaries used in this analysis.

\begin{table*}[htp]
\centering
\small
\caption{Residual region coverage after round-robin multi-driver fuzzing.
$\#Region$ denotes the number of structural regions in the shared backbone.
RC-Min, RC-Median, and RC-Max summarize the distribution of per-region coverage ratios ($rc$).
Frac($rc=0$), Frac($rc\leq0.2$), and Frac($rc\leq0.5$) report the fraction of regions that remain uncovered or weakly covered after fuzzing.}
\label{tab:rq5-residual-region-coverage}
\resizebox{0.90\textwidth}{!}{
\begin{tabular}{llrrrrrrr}
\toprule
Benchmark & Executable & $\#Region$ & RC-Min & RC-Median & RC-Max & Frac($rc=0$) & Frac($rc\leq0.2$) & Frac($rc\leq0.5$) \\
\midrule

\multirow{2}{*}{snort3}
 & snort & 87 & 0.0000 & 0.0000 & 0.6952 & 0.5517 & 0.8161 & 0.9655 \\
 & snort2lua & 24 & 0.1000 & 0.7042 & 1.0000 & 0.0000 & 0.0833 & 0.1667 \\

\multirow{2}{*}{http-parser}
 & parsertrace & 1 & 1.0000 & 1.0000 & 1.0000 & 0.0000 & 0.0000 & 0.0000 \\
 & url\_parser & 1 & 0.7143 & 0.7143 & 0.7143 & 0.0000 & 0.0000 & 0.0000 \\

\multirow{2}{*}{ffmpeg}
 & ffmpeg & 198 & 0.0000 & 0.0000 & 0.3333 & 0.8384 & 0.9798 & 1.0000 \\
 & ffprobe & 139 & 0.0000 & 0.0000 & 1.0000 & 0.6619 & 0.9281 & 0.9928 \\

\multirow{3}{*}{libtiff}
 & tiff2bw & 8 & 0.0000 & 0.4288 & 1.0000 & 0.1250 & 0.3750 & 0.8750 \\
 & tiffinfo & 6 & 0.0000 & 0.4207 & 0.5531 & 0.1667 & 0.1667 & 0.5000 \\
 & tiff2pdf & 8 & 0.0000 & 0.3472 & 0.7745 & 0.2500 & 0.2500 & 0.6250 \\

\multirow{3}{*}{wavpack}
 & wavpack & 10 & 0.0000 & 0.0000 & 0.1341 & 0.7000 & 1.0000 & 1.0000 \\
 & wvunpack & 8 & 0.0000 & 0.0000 & 0.2500 & 0.6250 & 0.8750 & 1.0000 \\
 & wvgain & 5 & 0.0000 & 0.0000 & 0.3409 & 0.8000 & 0.8000 & 1.0000 \\

\multirow{1}{*}{git}
 & git & 39 & 0.0000 & 0.0000 & 0.1270 & 0.7436 & 1.0000 & 1.0000 \\

\multirow{3}{*}{sleuthkit}
 & istat & 64 & 0.0000 & 0.0000 & 0.2016 & 0.8906 & 0.9844 & 1.0000 \\
 & img\_stat & 14 & 0.0000 & 0.6833 & 1.0000 & 0.1429 & 0.2143 & 0.2857 \\
 & tsk\_recover & 59 & 0.0000 & 0.0000 & 0.2366 & 0.8644 & 0.9831 & 1.0000 \\

\multirow{1}{*}{file}
 & file & 13 & 0.0000 & 0.6111 & 0.9091 & 0.0769 & 0.2308 & 0.3846 \\

\multirow{3}{*}{xpdf}
 & pdfdetach & 10 & 0.1500 & 0.4243 & 1.0000 & 0.0000 & 0.1000 & 0.6000 \\
 & pdfinfo & 10 & 0.1579 & 0.4127 & 0.8875 & 0.0000 & 0.1000 & 0.6000 \\
 & pdftops & 17 & 0.0000 & 0.4444 & 1.0000 & 0.0588 & 0.1765 & 0.5882 \\

\multirow{1}{*}{libxml2}
 & xmllint & 13 & 0.0000 & 0.2500 & 1.0000 & 0.2308 & 0.3846 & 0.6923 \\

\multirow{1}{*}{jq}
 & jq & 14 & 0.0949 & 0.7510 & 1.0000 & 0.0000 & 0.1429 & 0.2857 \\

\multirow{3}{*}{binutils}
 & objdump & 22 & 0.0000 & 0.1662 & 1.0000 & 0.3182 & 0.5000 & 0.8636 \\
 & readelf & 21 & 0.0000 & 0.0270 & 0.9231 & 0.4762 & 0.7143 & 0.8095 \\
 & addr2line & 16 & 0.0000 & 0.0913 & 1.0000 & 0.3750 & 0.8125 & 0.8750 \\

\multirow{1}{*}{cppcheck}
 & cppcheck & 89 & 0.0000 & 0.0000 & 1.0000 & 0.7416 & 0.9551 & 0.9775 \\

\multirow{1}{*}{libdwarf}
 & dwarfdump & 17 & 0.0000 & 0.1656 & 0.5641 & 0.2941 & 0.5294 & 0.8235 \\

\multirow{1}{*}{cpython3}
 & python & 33 & 0.0000 & 0.2439 & 1.0000 & 0.3030 & 0.4545 & 0.8788 \\

\multirow{2}{*}{quickjs}
 & qjs & 14 & 0.0000 & 0.4585 & 0.8535 & 0.0714 & 0.2143 & 0.5714 \\
 & qjsc & 13 & 0.0000 & 0.3361 & 0.8333 & 0.0769 & 0.3077 & 0.6154 \\

\multirow{1}{*}{lua}
 & lua & 10 & 0.7692 & 0.9393 & 1.0000 & 0.0000 & 0.0000 & 0.0000 \\

\multirow{2}{*}{libarchive}
 & bsdtar & 20 & 0.0000 & 0.0933 & 0.5000 & 0.4000 & 0.6500 & 1.0000 \\
 & bsdunzip & 9 & 0.3030 & 0.8235 & 1.0000 & 0.0000 & 0.0000 & 0.3333 \\

\multirow{1}{*}{upx}
 & upx & 27 & 0.0000 & 0.7918 & 1.0000 & 0.1111 & 0.2222 & 0.3333 \\

\multirow{1}{*}{xz}
 & xz & 14 & 0.0000 & 0.1804 & 0.5692 & 0.2143 & 0.5000 & 0.9286 \\

\multirow{3}{*}{hdf5}
 & h5dump & 10 & 0.0000 & 0.4002 & 0.7500 & 0.1000 & 0.2000 & 0.7000 \\
 & h5ls & 9 & 0.1765 & 0.3846 & 0.6573 & 0.0000 & 0.1111 & 0.7778 \\
 & h5repack & 8 & 0.0000 & 0.5328 & 0.7090 & 0.1250 & 0.1250 & 0.5000 \\

\multirow{3}{*}{netcdf}
 & ncdump & 12 & 0.0000 & 0.3046 & 0.9130 & 0.0833 & 0.3333 & 0.6667 \\
 & ncgen & 10 & 0.0000 & 0.0000 & 0.3015 & 0.6000 & 0.9000 & 1.0000 \\
 & nccopy & 11 & 0.0000 & 0.0000 & 0.0817 & 0.9091 & 1.0000 & 1.0000 \\

\multirow{1}{*}{sqlite3}
 & sqlite3 & 7 & 0.0197 & 0.1596 & 0.2575 & 0.0000 & 0.7143 & 1.0000 \\

\multirow{1}{*}{unbound}
 & checkconf & 18 & 0.0000 & 0.0000 & 0.7333 & 0.6667 & 0.8333 & 0.9444 \\

\bottomrule
\end{tabular}}
\vspace{-10pt}
\end{table*}

\subsubsection{Pattern Summary and Category Construction}

Across the benchmark suite, residual region coverage spans a broad range.
Some executables retain severe structural gaps after fuzzing,
with median region coverage equal to zero and large fractions of regions remaining uncovered or only weakly covered.
For example, \texttt{ffmpeg}, \texttt{ffprobe}, \texttt{git}, \texttt{istat}, \texttt{tsk\_recover}, \texttt{cppcheck}, \texttt{ncgen}, \texttt{nccopy}, and \texttt{checkconf}
all show \textit{RC-Median} values of zero together with high \textit{Frac($rc=0$)} and \textit{Frac($rc\leq0.5$)} values.
In these executables, under-exploration is not limited to a small hard tail.
Instead, a substantial portion of the shared structural backbone remains untouched or only marginally exercised.
We group these executables into a category of \emph{backbone-dominated residual gaps}.
This category is meaningful because the dominant pattern is a large residual region mass rather than a few isolated hard regions.
Such profiles indicate that equal time sharing across the available drivers is insufficient to activate much of the executable's structural space,
suggesting stronger access constraints such as driver-entry mismatch, format-specific logic, deeper semantic preconditions, or subsystems that are only weakly exercised by the current seed and scheduling setup.

A second group of executables exhibits a different pattern.
These programs achieve nontrivial structural exploration,
but still leave a broad low-coverage tail across the backbone.
Examples include \texttt{objdump}, \texttt{readelf}, \texttt{addr2line}, \texttt{xmllint}, \texttt{python}, \texttt{xz}, \texttt{bsdtar}, \texttt{pdftops}, \texttt{pdfinfo}, and \texttt{pdfdetach}.
Their \textit{RC-Median} values are above zero, indicating that fuzzing reaches a meaningful portion of the structural space,
yet their \textit{Frac($rc\leq0.5$)} values remain substantial,
showing that many regions are still only weakly explored.
We group these executables into a category of \emph{partial exploration with a substantial residual tail}.
Unlike the first category, these programs do not exhibit broad structural failure.
However, unlike the strongest executables, their residual under-exploration is still structurally diffuse rather than localized.
This category therefore captures cases where round-robin fuzzing can reach the main or easier regions of the backbone,
but struggles to extend exploration more uniformly into structurally harder areas.
The remaining gaps in this regime are less suggestive of total driver inadequacy
and more suggestive of uneven exploration pressure across subsystems, modes, or deeper execution stages.

At the other extreme, some executables achieve relatively broad structural exploration and leave only limited residual pockets.
Examples include \texttt{lua}, \texttt{jq}, \texttt{snort2lua}, \texttt{bsdunzip}, \texttt{img\_stat}, \texttt{file}, and \texttt{upx}.
These executables generally show high \textit{RC-Median} values, low \textit{Frac($rc=0$)}, and comparatively small \textit{Frac($rc\leq0.5$)} values.
In these programs, residual under-exploration is no longer a backbone-wide phenomenon,
but is instead concentrated in a limited set of regions.
We group them into a category of \emph{broad exploration with localized residual pockets}.
This regime captures cases where it separates executables whose remaining gaps are narrow and concentrated
from those whose residual structure is still broad or dominant.
Such profiles indicate that round-robin fuzzing is already broadly effective at the structural level,
and that the remaining hard regions are more likely tied to localized behaviors such as rare handlers, specialized modes, infrequent semantic branches, or narrow validation chains.

Substantial residual heterogeneity also remains among executables from the same benchmark family. 
For example, 
\texttt{snort} and \texttt{snort2lua}, \texttt{istat} and \texttt{img\_stat}, \texttt{bsdtar} and \texttt{bsdunzip}, 
and \texttt{nccopy} and \texttt{ncdump} exhibit clearly different residual profiles 
even though they belong to the same suite. 
This suggests that residual structural gaps are driven more 
by executable-level characteristics than by the benchmark family as a whole. 
It also indicates that under-exploration depends not only on the broader software family, 
but also on how each executable presents configuration modes, 
entry behavior, and internal structural pathways.

\subsubsection{Case Studies by Residual-Gap Category}

For readability, 
all case-study figures use the same notation and visual encoding.
$R_i$ denotes a residual region identifier, 
$RC$ its region coverage ratio after round-robin multi-driver fuzzing, 
$S$ the region size, 
$U$ representative uncovered function(s) in that region, and 
$C$ representative covered anchor or neighboring function(s) linking the region to already explored code
(``--'' indicates no clear covered anchor). 
In the graphs, 
red nodes denote uncovered regions, 
orange nodes denote weakly covered regions, and 
edges indicate structural connectivity among residual regions and the explored backbone. 
The quantitative patterns above show that residual under-exploration 
does not form a single homogeneous tail. 
Instead, 
different executables exhibit distinct residual-gap regimes, 
suggesting different underlying causes. 
We therefore examine representative case studies 
for each category to explain why such regions persist after round-robin multi-driver fuzzing.

\begin{figure}[htp]
  \centering
  \includegraphics[width=0.67\textwidth]{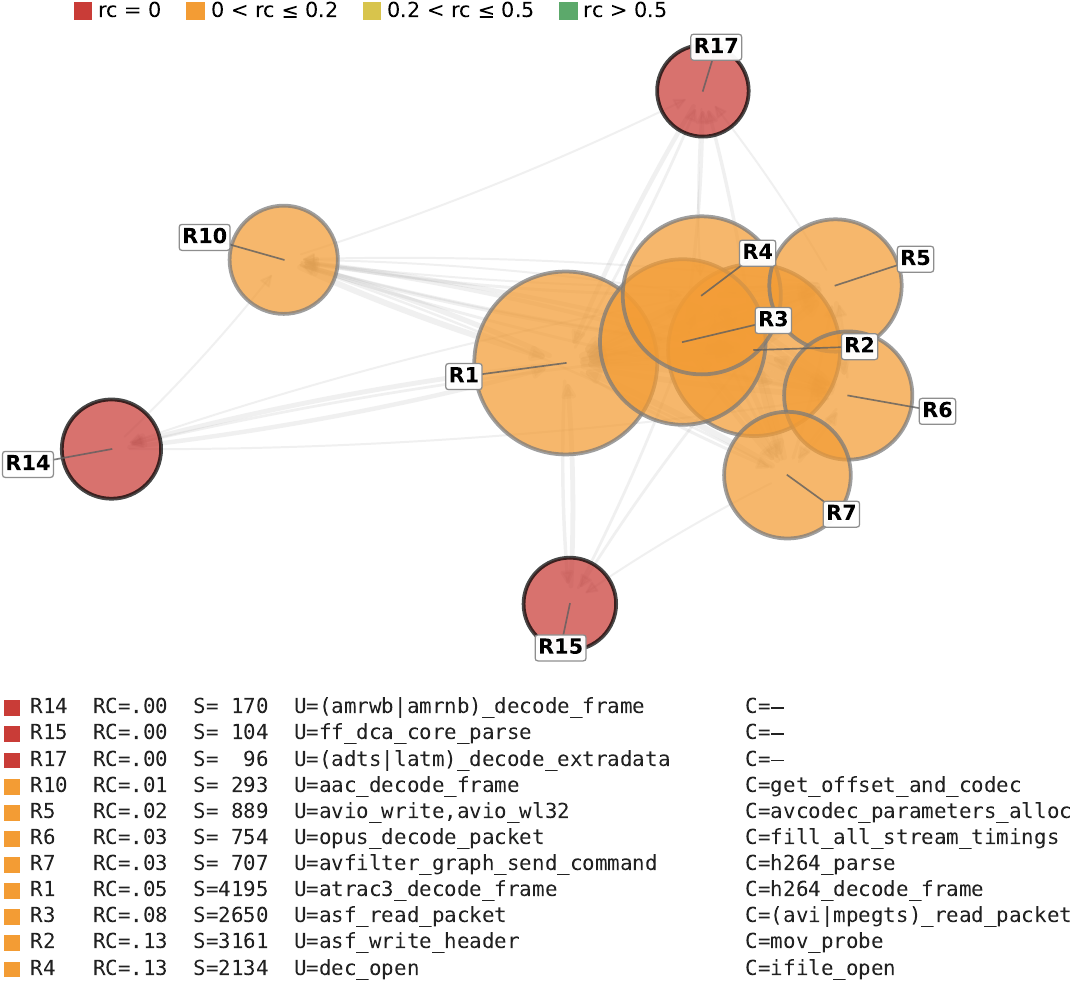}
  \caption{\textbf{Case Study A: Backbone-dominated residual gaps in \texttt{ffmpeg}.}
Several residual regions remain fully uncovered (e.g., R14, R15, and R17), while even the large backbone region R1 is only weakly exercised (RC=0.05). These regions span diverse codec- and format-specific handlers, indicating that the residual gap is broad and structurally connected rather than an isolated fringe, and persists because many paths in the shared decoding backbone require strict format- and state-dependent semantic conditions.}
  \label{fig:case-a}
  \vspace{-10pt}
\end{figure}

\vspace{3pt}
\noindent
\textbf{Case Study A: \texttt{ffmpeg} as a backbone-dominated residual-gap executable.}
\texttt{ffmpeg} is a representative example of the backbone-dominated residual-gap category.
At the executable level, 
it retains severe residual under-exploration after round-robin multi-driver fuzzing: 
among 198 structural regions, the median region coverage is 0, 
83.84\% of regions remain completely uncovered, and 97.98\% remain at or below a coverage ratio of 0.2. 
This profile indicates that the residual gap is not a small hard tail, but a backbone-wide phenomenon.

Figure~\ref{fig:case-a} helps explain why.
The uncovered and weakly covered regions are associated with diverse media-processing routines, 
including amrwb\_decode\_frame, aac\_decode\_parse, opus\_decode\_packet, h264\_parse, 
and atrac3\_decode\_frame.
These are not repetitions of one narrow subsystem.
Instead, 
they span multiple codec- and format-specific branches that sit behind distinct semantic preconditions, 
such as valid container structure, codec selection, stream metadata, packet layout, and stage-specific parser state.
Thus, 
even though round-robin scheduling exposes multiple drivers, 
equal time sharing alone does not make these branches easy to reach. 
The limiting factor is not merely driver availability, 
but the difficulty of constructing inputs that satisfy the right semantic gates for each specialized path.
The graph structure reinforces this interpretation.
Several residual regions remain completely uncovered and appear at the periphery of the explored backbone, while others are connected to already exercised regions but still receive only negligible coverage.
Most notably, R1 is large yet has RC=0.05, showing that the problem is not confined to tiny fringe components.
Even structurally important regions remain weakly explored.
This suggests that fuzzing can repeatedly enter the shared parsing and dispatch backbone, but tends to stay within easier or already-activated processing paths rather than crossing into harder codec-specific handlers.
In other words, the bottleneck is a semantic access barrier: the campaign reaches the neighborhood of these regions, but rarely satisfies the precise input conditions needed to traverse them.

Overall, this case study explains why \texttt{ffmpeg} falls into the backbone-dominated residual-gap category.
Its remaining gaps are broad, structurally connected, and semantically heterogeneous.
They persist because specialized decode and parse logic requires much stronger format-consistent inputs than round-robin scheduling and generic mutation alone can reliably provide.
This also clarifies why adding multiple drivers is still insufficient here: the main challenge is not just having more entry points, but generating inputs that can unlock the correct deep media-specific execution modes.

\vspace{3pt}
\noindent
\textbf{Case Study B: Partial exploration with a substantial residual tail.}
\texttt{xmllint} represents executables that achieve meaningful structural progress, 
yet still retain a broad low-coverage tail. 
Unlike the backbone-dominated failure pattern in Case Study~A, 
residual under-exploration here is uneven rather than uniform: 
several regions are already meaningfully explored, 
including DTD/external-subset parsing (R4, RC=0.76), serialization/output handling (R9, RC=0.55), and notation/encoding-related logic (R10, RC=0.51). 
However, 
a substantial residual tail remains across other structurally connected regions. 
In particular, 
XPath evaluation is entirely uncovered (R8, RC=0), 
shell/debug-style inspection logic is entirely uncovered (R11, RC=0), and 
schema-facet validation is also entirely uncovered (R12, RC=0). 
Other advanced behaviors remain only weakly explored, 
including schema validation (R2, RC=0.08), Relax NG validation (R3, RC=0.23), and
canonicalization (R6, RC=0.25). 
This pattern suggests that round-robin multi-driver fuzzing 
can reliably exercise the main document-processing backbone of \texttt{xmllint}, 
but does not extend coverage uniformly into specialized semantic modes. 
The remaining gaps persist not because the executable is broadly unreachable, 
but because deeper query, validation, and transformation behaviors depend on more specific option, document, 
and internal state combinations than generic round-robin fuzzing can consistently satisfy.

\begin{figure}[htp]
  \centering
  \includegraphics[width=0.65\textwidth]{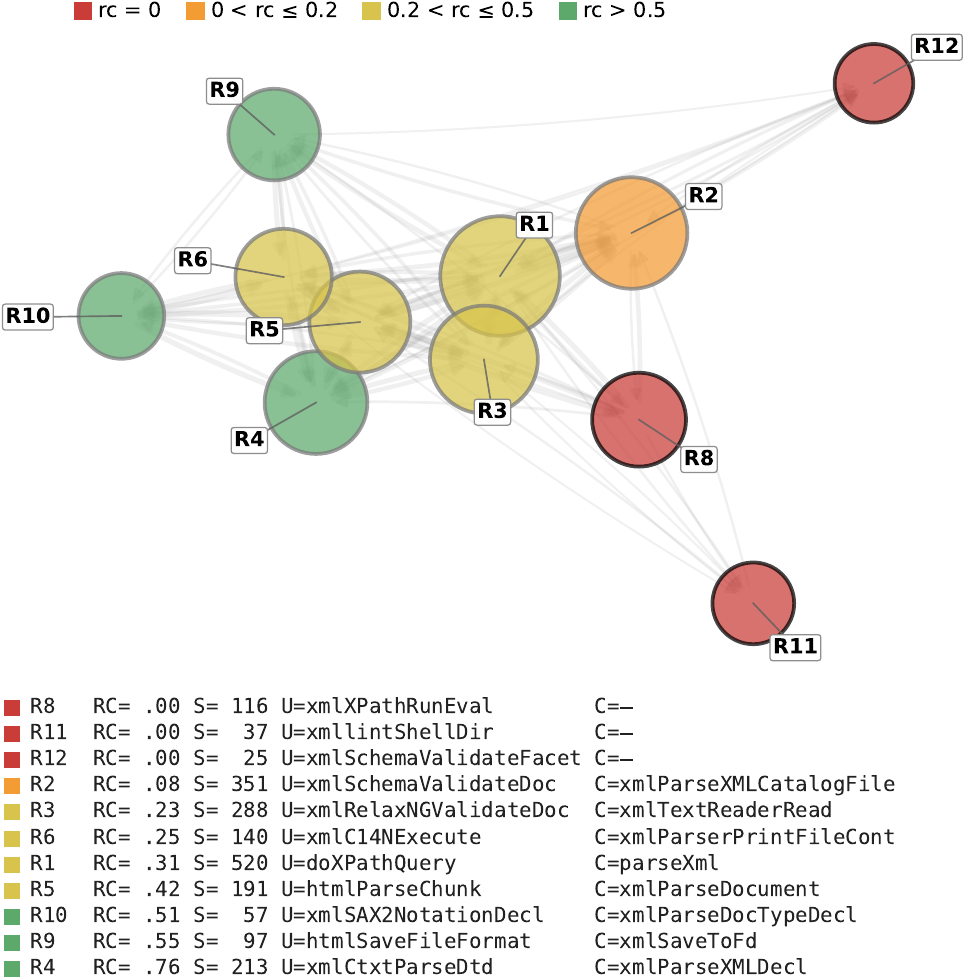}
  \caption{\textbf{Case Study B: Partial exploration with a substantial residual tail in \texttt{xmllint}.}
  \texttt{xmllint} shows a mixed residual profile: several regions are already
  meaningfully explored (e.g., R4, R9, and R10), but a substantial tail of
  structurally connected regions remains weakly explored or uncovered. The
  hardest residual gaps concentrate in semantically richer behaviors, including
  XPath evaluation (R8), shell/debug inspection (R11), schema-facet validation
  (R12), schema validation (R2), Relax NG validation (R3), and
  canonicalization (R6).}
  \label{fig:case-b}
  \vspace{-10pt}
\end{figure}

\vspace{3pt}
\noindent
\textbf{Case Study C: Broad exploration with localized residual pockets.}
\texttt{lua} represents the regime 
in which round-robin multi-driver fuzzing already achieves near-complete structural exploration, and 
the remaining residual regions are limited in number, well-connected, and only mildly underexplored.
Unlike the previous two categories, 
the residual profile here is not dominated by large uncovered backbone regions or a broad low-coverage tail. 
Instead, 
all shown residual regions are already substantially exercised, 
with coverage ratios ranging from 0.77 to 0.99. 
The remaining pockets concentrate in specific semantic behaviors, 
including pattern-matching and capture logic (R10), 
locale-sensitive numeric parsing (R6), script loading, 
and REPL-side execution (R2), 
binary chunk loading (R8), 
table/hash maintenance (R7), 
protected-call and coroutine recovery (R4), 
garbage-collection traversal of userdata and weak references (R5), VM/runtime edge cases (R3), and 
parser/code-generation diagnostics (R1). 
This pattern suggests that the main interpreter and runtime backbone of \texttt{lua} 
is already broadly reachable under multi-driver fuzzing, and 
that the remaining gaps persist 
mainly because these behaviors are triggered only under narrower semantic conditions, 
such as specific script constructs, malformed or boundary-case inputs, unusual runtime states, or
binary-chunk formats. 
In other words, 
the residual under-exploration is localized not because major functionality remains inaccessible, 
but because the remaining underexplored behaviors 
lie in specialized semantic corners of an otherwise well-explored executable.

\begin{figure}[t]
  \centering
  \includegraphics[width=0.67\textwidth]{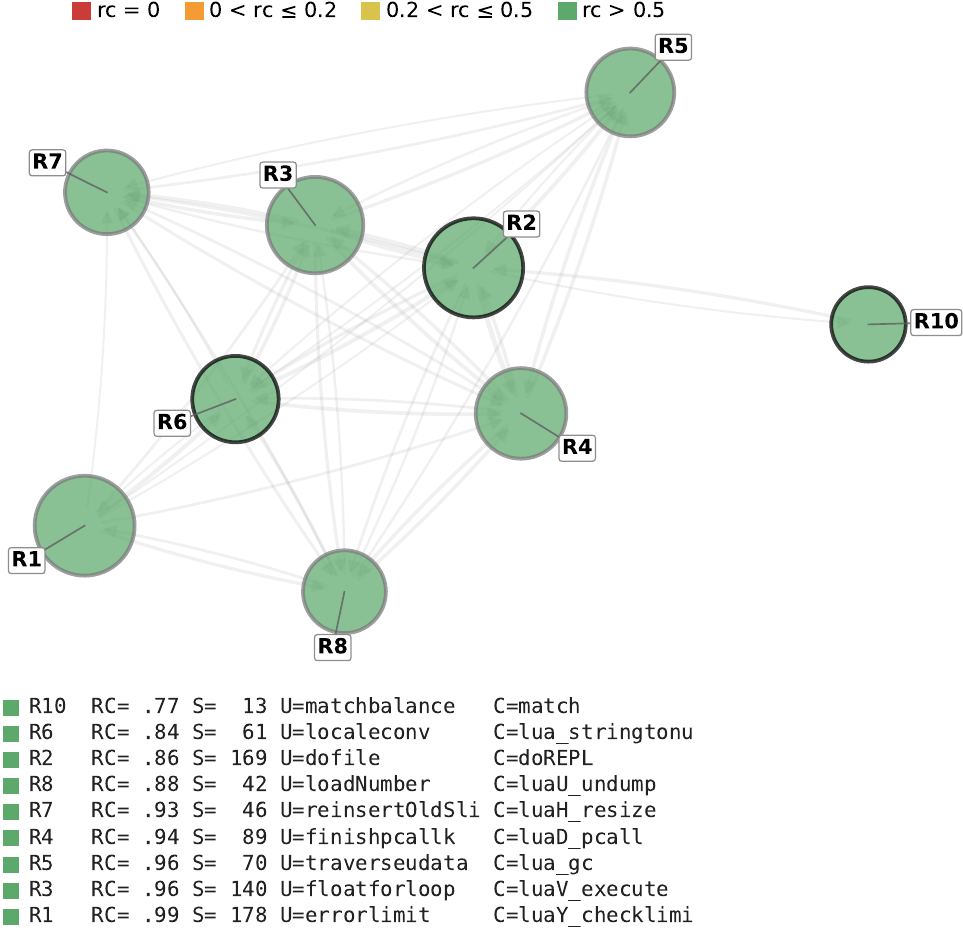}
  \caption{\textbf{Case Study C: Broad exploration with localized residual pockets in \texttt{lua}.}
  All shown residual regions are already highly explored (RC=0.77--0.99),
  indicating that round-robin multi-driver fuzzing covers most of the interpreter and runtime backbone.
  The remaining low-intensity pockets cluster in specialized semantic behaviors,
  including pattern matching, numeric parsing, script and binary-chunk loading, table/hash maintenance,
  protected-call recovery, garbage collection, VM edge cases, and parser errors.}
  \label{fig:case-c}
  \vspace{-10pt}
\end{figure}

\find{
Residual region coverage after round-robin multi-driver fuzzing remains highly heterogeneous 
across executables and falls into three recurring regimes: 
backbone-dominated gaps, partial exploration with a substantial residual tail, 
and broad exploration with only localized residual pockets. 
These regimes reflect different sources of persistent under-exploration, 
including stronger semantic barriers in some executables, 
uneven exploration of specialized paths in others, 
and only localized corner cases in the most thoroughly explored cases. 
Similar variation appears even within the same benchmark family, 
suggesting that residual structural difficulty is largely executable-specific rather than benchmark-wide. 
Overall, 
while multi-driver fuzzing substantially improves structural reachability, 
round-robin scheduling alone is often insufficient for uniform exploration of an executable’s shared backbone.
}

\section{Discussion} \label{sec:diss}

Our results show that multi-driver fuzzing is beneficial, 
but its effectiveness depends strongly on the structural relationships among drivers. 
RQ1 showed that multi-driver fuzzing generally improves structural exploration over the best single-driver baseline. 
RQ2 further showed that equal time allocation does not produce equal driver contribution: 
some drivers contribute disproportionately more than others, while some add little beyond already explored regions. 
RQ3 and RQ4 explain this behavior structurally, 
showing that driver-induced subgraphs differ not only in size, 
but also in cohesion, fragmentation, modular organization, and pairwise overlap. 
RQ5 extends this picture to residual structural coverage after combining drivers: 
executables do not exhibit one uniform residual tail, 
but instead fall into three recurring regimes: 
backbone-dominated residual gaps, partial exploration with a substantial residual tail, and broad exploration with localized residual pockets. Taken together, 
these results indicate that the benefit of multi-driver fuzzing is governed less by the mere number of drivers
than by the structural diversity and complementarity they introduce, 
and by how effectively fuzzing penetrates the remaining cold regions of the shared backbone.

A first implication is that driver sets should not be treated as flat collections of interchangeable execution modes.
Some executables, such as \texttt{sqlite3} and \texttt{upx}, exhibit highly redundant driver behavior:
drivers contribute uniformly, induce structurally similar subgraphs, and overlap heavily.
In such cases, adding more drivers yields little additional benefit.
By contrast, executables such as \texttt{ffmpeg}, \texttt{objdump}, \texttt{python}, and \texttt{lua}
show heterogeneous driver effectiveness, substantial structural variation, and mixed overlap structure.
Here, multiple drivers expose complementary regions of the program and can substantially improve exploration.
At the same time, RQ5 shows that such gains are not always sufficient to eliminate structural cold spots:
depending on the executable, the remaining gaps may still involve most of the shared backbone, a broad low-coverage tail, or only a few localized pockets.
This suggests that the value of a driver depends on its marginal structural contribution rather than its presence alone, and that multi-driver gains should be evaluated not only by total coverage improvement but also by how much residual cold structure remains.

A second implication is that equal scheduling is not an adequate general policy for multi-driver fuzzing.
Round-robin allocation is simple and fair at the driver level, but our results show that it does not ensure fairness at the structural level.
When multiple drivers repeatedly traverse overlapping structural cores,
time is spent reinforcing already hot regions, while cold regions remain under-explored.
RQ5 makes this effect more explicit by showing that residual under-exploration is not distributed uniformly:
in some executables, a large portion of the backbone remains cold even after combining multiple drivers,
whereas in others the remaining difficulty is concentrated in a smaller set of stubborn regions.
A scheduler that is unaware of structural redundancy may therefore waste fuzzing effort on drivers whose marginal contribution is low,
or fail to distinguish between executables that require broad reallocation of effort and those that require more targeted triggering of residual pockets.

More broadly, the results suggest that configuration-rich software should be viewed as a collection of partially overlapping execution subspaces.
Different drivers expose different slices of the same program, and the relationships among these slices shape fuzzing outcomes.
This perspective helps explain why multi-driver fuzzing sometimes yields dramatic gains and sometimes only marginal improvement.
It also helps explain why residual under-exploration persists in different forms across executables:
some programs remain blocked at a broad structural level, while others are limited mainly by a smaller set of semantically or structurally difficult regions.
Future fuzzing tools should therefore reason about driver structure explicitly,
rather than treating configuration modes as black-box entry points.

Our findings also have implications beyond scheduling.
For benchmark construction and evaluation, they suggest that reporting only aggregate coverage or bug counts may hide important structural differences among drivers.
Two executables may achieve similar total coverage while differing substantially in redundancy, fragmentation, or residual cold-region structure.
For tool builders, this means that structural analysis can provide actionable diagnostics:
it can identify redundant drivers, reveal complementary driver families, and expose persistently cold regions that may require targeted generation, specialized inputs, or different scheduling priorities.

Several limitations should also be noted.
Our structural analysis is based on driver-induced subgraphs projected onto a static call graph,
which provides a useful common reference but may not capture all dynamic relationships precisely.
The region-level under-exploration analysis identifies persistent cold structure,
but the quantitative summaries in RQ5 do not by themselves explain why a region remains hard to reach.
That question requires follow-up inspection, which is why we complement the quantitative categorization with category-based case studies.
In addition, the current study focuses on round-robin multi-driver fuzzing rather than more advanced adaptive scheduling strategies.
These limitations do not affect the central findings,
but they define natural directions for future work.

Overall, the main lesson is that multi-driver fuzzing is fundamentally a structural exploration problem.
Its success depends on how drivers partition, overlap, and complement one another in the program graph,
and on how effectively fuzzing penetrates the residual cold structure that remains after combining them.
Effective multi-driver fuzzing therefore requires more than simply adding drivers:
it requires selecting, organizing, and scheduling them based on their structural relationships and their marginal contribution to under-explored regions.

\subsection{Practical Implications}

Our results suggest several practical implications for the design of driver-aware fuzzers.

\vspace{3pt}
\noindent
\textbf{(1) Measure the marginal value of each driver.}
Different drivers contribute unevenly to exploration.
A driver should not be treated as useful simply because it is present in the pool.
Instead, a fuzzer should estimate how much new behavior or previously unreached program regions each driver contributes beyond what other drivers have already exposed.
This implication follows directly from the observed heterogeneity in driver effectiveness across executables.

\vspace{3pt}
\noindent
\textbf{(2) Account for overlap and redundancy among drivers.}
Our overlap analysis shows that many drivers repeatedly exercise highly shared program logic, especially common backbone behavior.
This suggests that driver-aware fuzzers should explicitly reason about overlap, for example by grouping similar drivers, reducing effort on highly redundant ones, or selecting a smaller but more diverse driver set.
In other words, adding more drivers is not always beneficial if they largely revisit the same behavior.

\vspace{3pt}
\noindent
\textbf{(3) Prioritize drivers that extend exploration into weakly covered regions.}
Some drivers continue to push exploration into weakly covered parts of the program, while others mainly revisit already hot paths.
This suggests that driver prioritization should depend not only on execution frequency or aggregate coverage gain, but also on whether a driver continues to open access to colder regions.
Drivers that keep extending exploration into such regions are likely to be more valuable than drivers that mainly reinforce already saturated behavior.

\vspace{3pt}
\noindent
\textbf{(4) Reallocate effort differently for different residual regimes.}
Our uncovered-region analysis reveals several recurring residual regimes, including backbone-dominated gaps, substantial residual tails, and localized residual pockets.
These patterns likely reflect different forms of remaining difficulty and therefore should not be treated in the same way.
Broad residual tails may call for broader reallocation toward under-explored modes, whereas localized pockets may benefit more from targeted triggering of narrow corner cases.
This suggests that remaining gaps should be interpreted by regime when deciding how to redirect fuzzing effort.

\vspace{3pt}
\noindent
\textbf{(5) Use richer feedback than aggregate coverage alone.}
Conventional feedback such as edge or function coverage summarizes how much of the program has been explored, but it does not reveal which regions are repeatedly revisited and which remain persistently cold.
Our results therefore suggest that effective multi-driver fuzzing should incorporate finer-grained feedback in addition to aggregate coverage metrics.
Such feedback can better reveal whether fuzzing is broadening exploration or merely reinforcing an already hot shared core.

\vspace{3pt}
\noindent
\textbf{(6) Improve driver--input combinations for persistent residual regions.}
Some regions remain under-explored even after combining multiple drivers.
At the same time, the current study does not disentangle whether these gaps arise primarily from driver choice, input quality, or their interaction.
A more defensible next step is therefore not to assume a driver-only or seed-only cause, but to improve the driver--input combinations associated with persistent residual regions.
This may include better driver organization, more targeted invocation patterns, specialized input support, or other targeted exploration mechanisms for the remaining gaps.

Overall, these implications show that effective multi-driver fuzzing requires more than simply running many drivers together.
It requires reasoning about the distinct contribution of each driver, the redundancy among drivers, and the different forms of residual under-exploration that remain after combining them.
\section{Threats to Validity} \label{sec:threat}

\textbf{Internal validity.}
A threat to internal validity is whether the observed differences across drivers and executables are caused by the proposed analysis setting rather than by uncontrolled experimental factors.
To reduce this risk, all comparisons use the same benchmark suite, the same driver specifications, and the same total fuzzing budget.
For RQ1, multi-driver fuzzing is compared against the best single-driver baseline under equivalent budget.
For RQ2--RQ5, all structural analyses are derived from dynamically observed execution results and are projected onto a shared context-insensitive static call graph.
Nevertheless, fuzzing is inherently stochastic, and these structural measurements depend on the executions reached during fuzzing.
Randomness in seed selection, mutation outcomes, path discovery, and scheduling decisions may alter the covered nodes and edges, and therefore also affect the resulting driver-induced subgraphs and their derived statistics.
As a result, changes in the initial seed corpus or in fuzzing randomness could lead to different observed structural measurements, including the residual region-coverage profiles used in RQ5.
Our results should therefore be interpreted as characterizing structural behavior under the studied fuzzing configuration rather than as invariant properties of the programs themselves.

\noindent
\textbf{Construct validity.}
A threat to construct validity is whether the selected metrics and constructed driver sets adequately capture the intended notions of effectiveness, heterogeneity, redundancy, and under-exploration.
We use covered call-graph nodes and control-flow graph edges as proxies for structural exploration in RQ1,
normalized per-driver call-graph coverage shares for driver effectiveness in RQ2,
LCC ratio, \#WCC, and modularity for structural organization in RQ3,
pairwise IoU for overlap and redundancy in RQ4,
and region-level coverage summaries for under-exploration in RQ5.
In particular, RQ5 uses per-region coverage ratios together with summary statistics such as RC-Min, RC-Median, RC-Max, and threshold-based fractions of weakly covered regions, e.g., Frac($rc=0$), Frac($rc\leq0.2$), and Frac($rc\leq0.5$), to characterize the severity and shape of residual structural gaps.
These metrics capture complementary structural aspects, but they are derived from dynamically observed execution results rather than exhaustive program behavior.
Moreover, the analyses are projected onto a context-insensitive static call graph.
Such a graph provides a common structural reference across drivers and executables,
but it may over-approximate feasible caller--callee relationships and may not precisely distinguish context-specific call behavior.
As a result, some measured structural properties, such as connectivity, modularity, overlap, and residual region structure,
may differ from those obtained under a more precise context-sensitive graph model.
In addition, the structural regions used in RQ5 are analysis units induced by community detection on the shared call-graph backbone.
Although this provides a systematic way to localize residual gaps, different graph partitioning choices could produce somewhat different region boundaries and therefore somewhat different residual-coverage summaries or category assignments.
The category construction in RQ5 should therefore be understood as a useful structural abstraction for organizing persistent under-exploration, rather than as a unique decomposition of the program.
In addition, the analyzed drivers are constructed from main configuration options only.
They do not exhaustively represent finer-grained parameter settings or other execution-shaping factors such as environment variables, configuration files, or runtime settings.
The conclusions should therefore be understood as characterizing the studied driver abstraction
and the observed structural exploration under a context-insensitive graph model,
rather than the full semantic configuration space or exact dynamic calling structure of each program.

\noindent
\textbf{External validity.}
A threat to external validity is whether the findings generalize beyond the evaluated benchmarks, constructed drivers, and fuzzing configuration.
Our study covers a diverse set of widely used configuration-rich programs from domains including media processing, toolchains, language runtimes, storage systems, and document processing.
This diversity improves generalizability, and several patterns recur across domains.
However, the results may not directly transfer to software with very different architectures, to programs with limited configuration diversity, or to ecosystems where driver behavior is not well represented by command-line modes.
In addition, the driver sets used in this study are constructed from main configuration options rather than from an exhaustive enumeration of all possible execution modes.
We do not systematically vary option parameters within each driver,
nor do we model other factors that may influence execution behavior,
such as environment variables, configuration files, runtime settings, or external system state.
Driver generation therefore remains only a partial approximation of the full configuration space,
and driver generation itself remains an open research problem.
Different driver-construction strategies may expose different structural regions, overlap patterns, effectiveness distributions, and residual region-coverage regimes.
Similarly, the findings are based on the studied fuzzing setting and may differ under alternative seed corpora, input models, region-construction choices, or adaptive scheduling strategies.
Thus, while the observed structural patterns appear broad, they should not be interpreted as universal across all programs, configuration mechanisms, driver-generation methods, or fuzzing scenarios.

\noindent
\textbf{Conclusion validity.}
A threat to conclusion validity is whether the empirical evidence is sufficient to support the claimed relationships among driver diversity, structural overlap, and fuzzing outcomes.
To mitigate this risk, we analyze the problem from multiple complementary perspectives:
overall effectiveness (RQ1), contribution imbalance (RQ2), graph organization (RQ3), overlap and redundancy (RQ4), and structural under-exploration (RQ5).
The consistency of these results strengthens the overall conclusions.
Even so, some interpretations remain inferential.
For example, we argue that overlap can reinforce already-hot regions, that structural complementarity can improve multi-driver gains,
and that different residual-gap regimes in RQ5 reflect different kinds of remaining exploration difficulty.
These interpretations are supported by converging structural evidence and category-based case studies, but they are not established through controlled causal manipulation.
Therefore, the paper supports strong empirical associations and design implications,
but causal claims about the exact mechanisms of driver interaction or residual under-exploration should be interpreted with appropriate caution.
\section{Conclusion}\label{sec:conclusion}

This paper studied how multi-driver fuzzing explores software structure in
real-world configuration-rich programs.
We introduced an execution-driven structural abstraction that represents each
program with a shared function-level call graph and derives driver-induced
subgraphs from driver-specific dynamic executions.
Using this abstraction, we conducted a large-scale empirical study over
OSS-Fuzz-derived benchmarks to analyze structural diversity, overlap,
imbalance, and under-exploration across execution modes.
Our results show that multi-driver fuzzing generally improves coverage over
the best single-driver baseline, but also exhibits strong imbalance across
drivers and leaves substantial structural regions persistently under-explored.
These findings indicate that multi-driver fuzzing is not merely a matter of
running more drivers, but of understanding how different execution modes
partition and complement the program graph.
Overall, this work provides an empirical foundation for more structure-aware
fuzzing.
Future work can build on this perspective to design better driver selection,
scheduling, and feedback strategies for multi-mode software.

\bibliographystyle{ACM-Reference-Format}
\bibliography{reference}

\appendixcontent

\end{document}